\documentclass[preprint2]{aastex}
\usepackage{epsf,graphicx}

\begin{document}

\def\la{\mathrel{\hbox{\rlap{\hbox{\lower4pt\hbox{$\sim$}}}\hbox{$<$}}}}
\def\ga{\mathrel{\hbox{\rlap{\hbox{\lower4pt\hbox{$\sim$}}}\hbox{$>$}}}}

\font\sevenrm=cmr7
\def\MgII{Mg~{\sevenrm II}}

\newdimen\digitwidth
\setbox0=\hbox{-.}
\digitwidth=\wd0
\catcode `@=\active
\def@{\kern\digitwidth}

\title{A Search for Synchrotron X-ray Emission in Radio Quasars}

\author{Hermine Landt}
\affil{Harvard-Smithsonian Center for Astrophysics, 60 Garden Street, 
Cambridge, MA 02138.}

\author{Paolo Padovani}
\affil{European Southern Observatory, Karl-Schwarzschild-Str. 2,
D-85748 Garching, Germany.}

\author{Paolo Giommi, Matteo Perri}
\affil{ASI Science Data Center, c/o ESRIN, via G. Galilei,
I-00044 Frascati, Italy.}

\author{Chi C. Cheung\altaffilmark{1}}
\affil{Kavli Institute for Particle Astrophysics and Cosmology,
Stanford University, Stanford, CA 94305.}

\altaffiltext{1}{Jansky Postdoctoral Fellow, National Radio Astronomy
Observatory}

\begin{abstract}

This paper presents {\sl XMM-Newton} and {\sl Chandra} X-ray
spectroscopy of ten flat-spectrum radio quasars (FSRQ) which are
candidates to have an X-ray spectrum dominated by jet synchrotron
emission. In all these FSRQ, which are less strongly relativistically
beamed than blazars, a considerable contribution from a power-law
component similar to that present in radio-quiet quasars is required
to explain the observed X-ray fluxes and X-ray spectral slopes. And as
in radio-quiet quasars, their relatively high optical/UV fluxes can be
accounted for by a significant contribution from thermal accretion
disk emission. The lack of success in finding radio quasars with
synchrotron X-rays is attributed to the adopted selection criteria,
which were based on the multiwavelength flux ratios of BL Lacertae (BL
Lac) objects. A refined selection technique, which additionally
involves radio imaging, is proposed to search for these important
candidates for the Gamma Ray Large Area Space Telescope (GLAST). On
the other hand, the discovered FSRQ with their strong accretion disk
signatures are expected to be important probes for studies of the
poorly known accretion disk - jet connection.

\end{abstract}


\section{Introduction}

Blazars offer us the unique possibility to study the spectral energy
distributions (SEDs) of relativistic jets. Being radio galaxies with
their jets oriented at relatively small angles with respect to our
line of sight they profit from the relativistic beaming effect. This
enhances dramatically (by factors up to $\sim 1000$) the intrinsic jet
(core) flux and thus makes it the dominant continuum emission at all
frequencies [see \citet{Urry95} for a review].

As first discussed by \citet{Jones74} two emission processes dominate
the multiwavelength SEDs of relativistic jets and, therefore, of
blazars, namely, synchrotron and inverse Compton. These give rise to
two prominent emission peaks in a logarithmic $\nu f_\nu - \nu$ plot,
which were observationally demonstrated by \citet{Sam96} and
\citet{Fos98}. The synchrotron emission peak ($\nu_{\rm peak}$) in
blazars is observed within a wide range of frequencies (from infrared
to soft-X-rays) in sources with weak emission lines, the BL Lacertae
objects \citep[BL Lacs; e.g.,][]{Gio95, Fos98, Gio05}. However,
strong-lined blazars, the flat-spectrum radio quasars (FSRQ), had
until recently synchrotron emission peaks at relatively low energies
and thus X-rays dominated by inverse Compton emission
\citep[e.g.,][]{Gam03, Tav02, Sam06}.

Based on this finding some authors proposed the so-called ``blazar
sequence'' \citep{Sam96, Fos98, Ghi98}. This scenario advocates that
the frequency of the synchrotron emission peak is governed by particle
Compton cooling by an external radiation field. Such a field, produced
by, e.g., the accretion disk or broad emission line region (BLR), both
more luminous in radio quasars than in BL Lacs, is expected to
interact with the particles in the jet via the inverse Compton process
causing them to lose energy. Therefore, within this scenario radio
quasars with $\nu_{\rm peak} > 10^{15}$ Hz and thus X-rays dominated
by synchrotron emission are not expected to exist.

Two new surveys have recently shown that mere selection effects had
prevented us so far from finding ``X-ray loud'' radio quasars,
allowing for the possibility that radio quasars with high $\nu_{\rm
peak}$ existed. About 10\% of the FSRQ discovered in the Deep X-ray
Radio Blazar Survey \citep[DXRBS;][]{Per98, L01} and $\sim 30\%$ of
the ones identified in the {\it ROSAT} All-Sky Survey (RASS) - Green
Bank \citep[RGB;][]{LM98, LM99} have multiwavelength flux ratios
similar to those of BL Lacs with synchrotron X-rays \citep{P03}. In
addition, subsequent investigations of the {\it EINSTEIN} Medium
Sensitivity Survey (EMSS) and the Slew survey revealed that these
radio quasars had gone undetected in previous surveys \citep{Wol01b,
Per01}.

The definite proof of the synchrotron nature of the X-ray emission in
these newly discovered blazars, however, requires X-ray
spectroscopy. \citet{P02} presented {\sl Beppo}SAX observations of
four ``X-ray loud'' FSRQ and found that the synchrotron emission of
one source, RGB J1629$+$4008, peaked at UV frequencies ($\nu_{\rm
peak} \sim 2\times10^{16}$ Hz) and dominated the X-rays. In this paper
we present {\sl XMM-Newton} and {\sl Chandra} observations of a
further ten of these sources.

The observed sample was selected as discussed in Section 2 and the
data acquired and analyzed as described in Section 3. The
multiwavelength SEDs of the sample are presented in Section 4 and the
results discussed in Section 5. Section 6 gives a brief
summary. Throughout this paper cosmological parameters $H_0 = 70$ km
s$^{-1}$ Mpc$^{-1}$, $\Omega_m=0.3$, and $\Omega_{\Lambda}=0.7$ have
been assumed. Energy spectral indices have been defined as $f_\nu
\propto \nu^{-\alpha}$ and photon spectral indices as $N(E) \propto
E^{-\Gamma}$, where $\Gamma=\alpha+1$.

\section{The Sample Selection}

We have selected for X-ray spectroscopy FSRQ from the DXRBS and RGB
survey which had multiwavelength flux ratios similar to those of BL
Lacs with X-rays dominated by synchrotron emission. To this selection
criterion we have added different constraints in the case of the {\sl
XMM-Newton} and the {\sl Chandra} observed samples (see below).

\begin{figure*}
\centerline{
\includegraphics[scale=0.38]{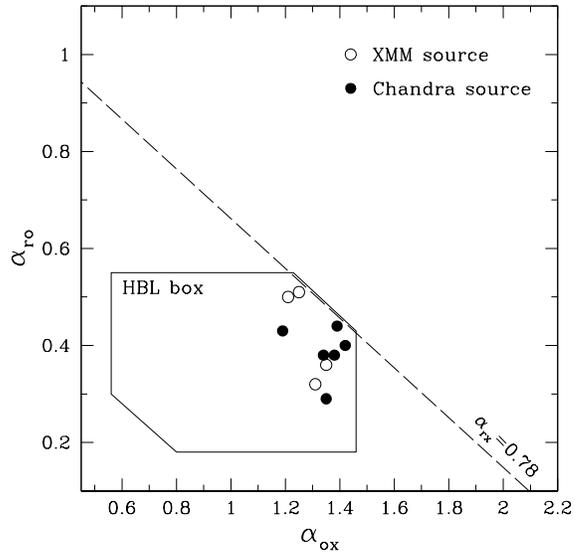}
}
\caption{\label{aroaox} ($\alpha_{\rm ro}, \alpha_{\rm ox}$) plane for
FSRQ from the DXRBS and RGB survey observed with {\sl XMM-Newton}
(open circles) and {\sl Chandra} (filled circles). The effective
spectral indices are defined in the usual way and calculated between
the rest-frame frequencies of 5 GHz, 5000~\AA, and 1 keV. The dashed
line represents the locus of constant $\alpha_{\rm rx}=0.78$, with
$\alpha_{\rm rx} \ga 0.78$ and $\alpha_{\rm rx} \la 0.78$ being
typical of BL Lacs with X-rays dominated by inverse Compton (LBL) and
synchrotron radiation (HBL), respectively. The ``HBL box'' as defined
by \citet{P03} represents the 2 $\sigma$ region around the mean
$\alpha_{\rm ro}, \alpha_{\rm ox}$, and $\alpha_{\rm rx}$ values of
HBL.}
\end{figure*}

As first suggested by \citet{P95} BL Lacs that emit synchrotron
radiation which peaks at low (IR/optical) frequencies, the so-called
low energy-peaked BL Lacs (LBL), can be distinguished from those that
emit synchrotron radiation which peaks at higher (UV/soft-X-ray)
frequencies, the so-called high energy-peaked BL Lacs (HBL), by their
distinct ratios between their radio, optical and X-ray fluxes. These
ratios are usually studied by plotting the objects in an ($\alpha_{\rm
ro}, \alpha_{\rm ox}$) plane, where $\alpha_{\rm ro}$ and $\alpha_{\rm
ox}$ are the usual rest-frame effective spectral indices defined
between 5 GHz, 5000~\AA, and 1 keV. In this plane constant
$\alpha_{\rm rx}$ values are represented by straight lines of slope
$\sim -0.5$ (see Fig. \ref{aroaox}).

A sequence of synchrotron emission peak frequencies in BL Lacs
produces a specific trail in the ($\alpha_{\rm ro}, \alpha_{\rm ox}$)
plane \citep{P95}. While the synchrotron emission peak is at low
enough frequencies for the X-rays to be dominated by inverse Compton
radiation (as in LBL), $\alpha_{\rm rx}$ is roughly constant, whereas
$\alpha_{\rm ro}$ decreases and $\alpha_{\rm ox}$ increases as the
peak moves to higher energies (see also \citet{Gio02}). When the
synchrotron radiation starts to dominate the X-ray band (as in HBL),
$\alpha_{\rm rx}$ starts to decrease, with LBL and HBL having values
of $\alpha_{\rm rx} > 0.78$ and $\le 0.78$, respectively
\citep{P96}. Now $\alpha_{\rm ro}$ is roughly constant (and low) and
$\alpha_{\rm ox}$ decreases. Based on these considerations \citet{P03}
have defined a region in the ($\alpha_{\rm ro}, \alpha_{\rm ox}$)
plane expected to be populated by high-energy peaked blazars, both BL
Lacs and FSRQ (see their Fig. 1). This so-called ``HBL box''
represents the 2 $\sigma$ region around the mean $\alpha_{\rm ro},
\alpha_{\rm ox}$, and $\alpha_{\rm rx}$ values of all HBL in the
multifrequency active galactic nuclei (AGN) catalog of \citet{P97}.

We have selected for X-ray spectroscopy with {\sl XMM-Newton} and {\sl
Chandra} FSRQ that populate the ``HBL box'' (Fig. \ref{aroaox}). For
observations with {\sl XMM-Newton} we have chosen FSRQ from DXRBS only
and have additionally required that the synchrotron emission peak
frequency, estimated from a crude multiwavelength SED including only
radio, optical and X-ray data points \citep{P03}, was $\nu_{\rm
peak}>10^{15}$ Hz. Eight sources satisfied these criteria and were
proposed. We were granted observing time in Cycle 3 for 4/8 sources
(PI: Padovani). For observations with {\sl Chandra} we have selected
sources from the sample of DXRBS and RGB FSRQ imaged with the VLA by
\citet{L06} and made an additional constraint that extended radio
structure was present. In this case the goal was to study the extended
X-ray morphology in addition to the core spectrum. Six sources
satisfied these criteria and were proposed. We were granted observing
time in Cycle 6 for all six sources (PI: Landt).

Table \ref{general} summarizes the general properties of our
sample. For all ten X-ray observed sources we have deep radio
observations obtained with the VLA at 1.4 GHz in A configuration and
for 8/10 sources also in C configuration. These data have been
published recently \citep{L06} and we reproduce here for convenience
also some of these radio measurements.

\section{The X-ray Observations}

On-board {\sl XMM-Newton} we used the European Photon Imaging Camera
(EPIC). In Table \ref{xmm} we give the journal of observations for the
three CCD cameras, MOS 1, MOS 2 and PN. All three cameras were
operated in small window mode and the thin filter was used. The source
WGA J1026$+$6746 was observed twice. However, the first observation
taken on 23 March 2004 was strongly affected by high radiation and,
therefore, not useful for our purposes.

On-board {\sl Chandra} we used the Advanced CCD Imaging Spectrometer
(ACIS) with the back-illuminated S3 chip at the aim-point. All sources
were observed with a 1/8 subarray, except for WGA J2347$+$0852 which
was observed with a 1/4 subarray. The observations were performed in
faint timed mode. In order to avoid that the readout streak falls on
the extended jet structures of our sources we additionally imposed
roll angle constraints. In Table \ref{chandra} we give the journal of
observations.

\subsection{The X-ray Data Analysis}

The EPIC data were processed using the XMM Science Analysis Software
(SAS; version 6.1.0). The initial data files were reprocessed with the
EMPROC and EPPROC scripts with default settings, using the latest
known calibration files (as of August 2005). We used X-ray events with
patterns $0-12$ and energies in the range $0.2 - 10$ keV for the two
MOS cameras. For the PN, we selected X-ray events with patterns $0-4$
(single and double pixel events) and energies in the range $0.2-15$
keV. The non-X-ray background was relatively high only in the case of
WGA J0110.5$-$1647. For this source we excluded from the analysis of
the PN in the energy range $>10$ keV time intervals with count rates
$\ge1$ counts s$^{-1}$. Source spectra were extracted from circular
regions of $30''$ radius for the two MOS and $35''$ radius for the
PN. This corresponds to an encircled energy of $\sim 85\%$. Background
spectra were taken from a similar circular region, offset from the
source position. None of the source light curves showed variability
over the duration of the observations after the removal of the
background.

The ACIS initial event files were reprocessed using the CIAO software
package (version 3.3) applying the latest known calibration files (as
of February 2006). We filtered for bad grades and used only X-ray
events with energies in the range $0.3-10$ keV. High non-X-ray
background periods were not present in our data. Source spectra were
extracted from circular regions of $3''$ radius, corresponding to an
encircled energy of $>95\%$ at 1.5 keV. Background spectra were taken
from a similar circular region, offset from the source position. None
of the source light curves showed variability over the duration of the
observations after the removal of the background.

\subsection{The X-ray Spectral Fits}

The background subtracted EPIC and ACIS spectra were fit using XSPEC
(version 12.2), with the individual response and ancillary matrices
produced with SAS and CIAO, respectively, from the source
spectra. Spectra were binned to a minimum of 20 counts per bin in
order to apply the $\chi^2$ minimization technique.

We initially fit the individual X-ray spectra from the three EPIC
detectors in order to check for consistency between the datasets. From
a simple power-law fit we found that the photon indices agreed within
$2\sigma$ and that the relative normalizations were consistent to
within 10\%. Therefore, for all sources we proceeded to fit the three
datasets simultaneously, allowing for the individual normalizations to
vary.

We fit the data both with single and broken power-law models with
photo-electric absorption using Wisconsin cross-sections from
\citet{Morr83}. The hydrogen column densities $N_{\rm H}$ were fixed
to the Galactic value \citep{DL90} and in the case of the single
power-law fits also allowed to vary in order to check for internal
absorption and/or indications of a ``soft excess''. The results for
the single power-law fits to the EPIC and ACIS spectra are reported in
Tables \ref{epic1} and \ref{acis1}, respectively. The results for the
broken power-law fits to the EPIC and ACIS spectra are reported in
Tables \ref{epic2} and \ref{acis2}, respectively.

\subsection{Discussion of Individual Sources} \label{discuss}

\begin{figure*}
\centerline{
\includegraphics[scale=0.34,angle=-90]{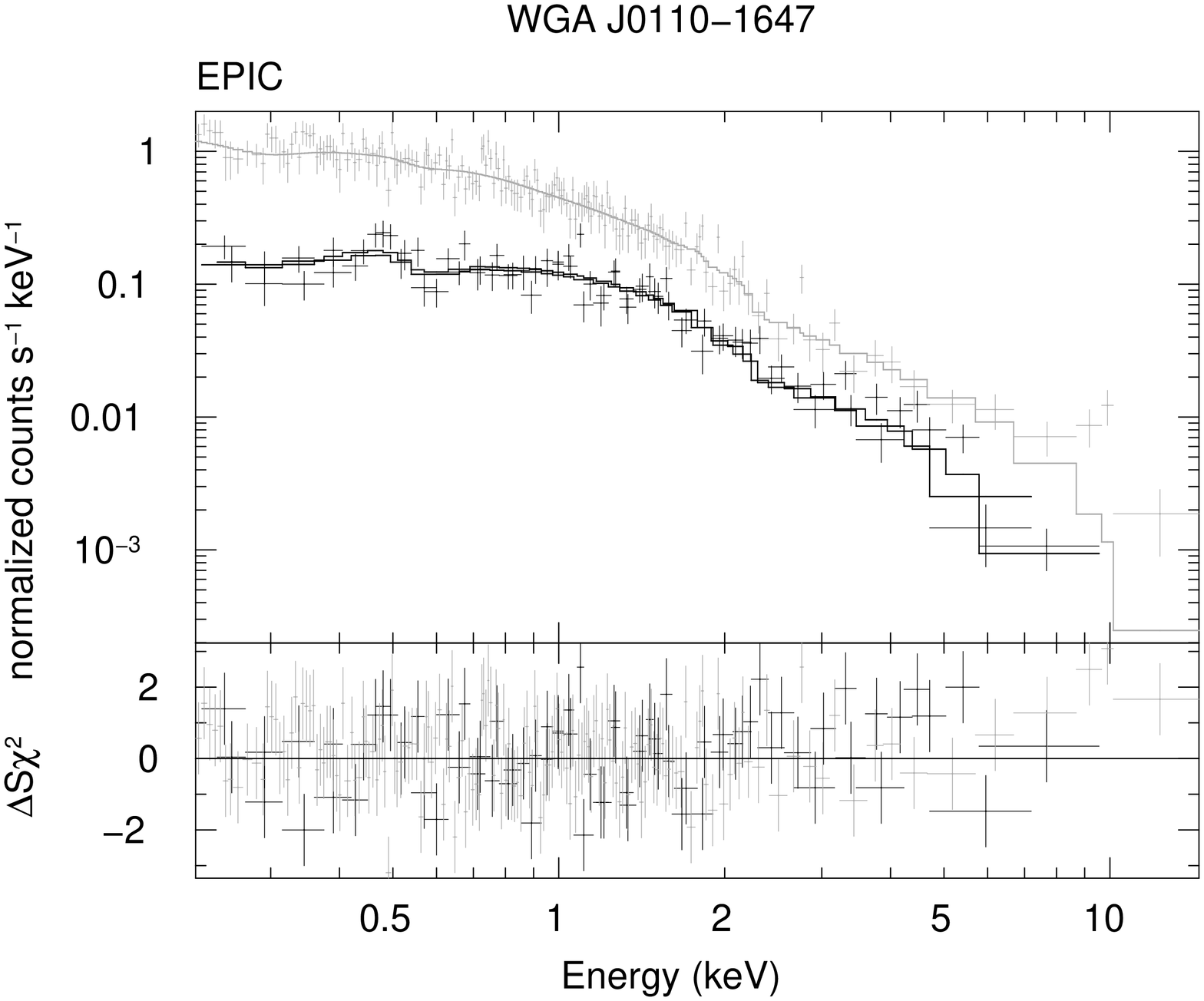}
\includegraphics[scale=0.34,angle=-90]{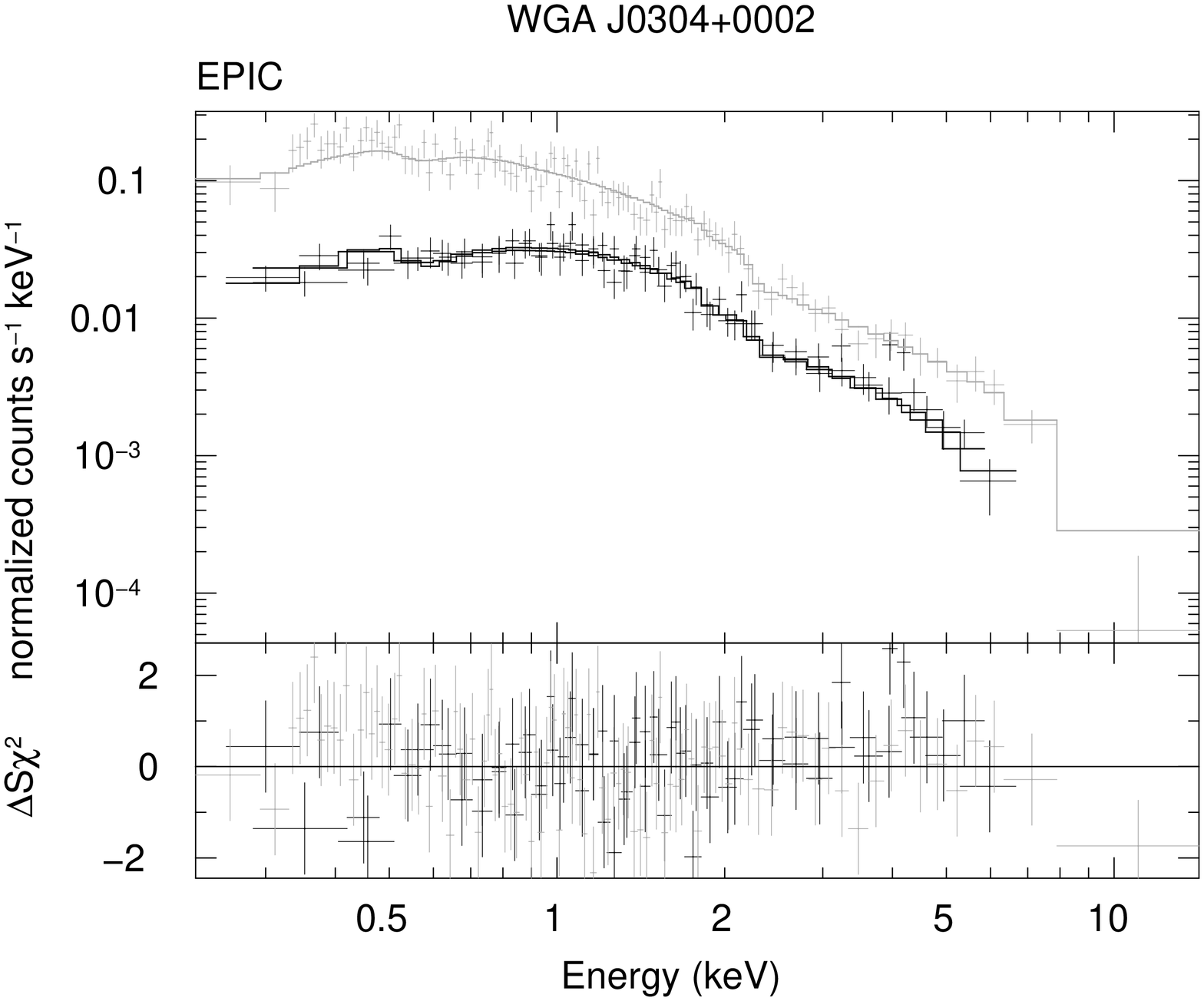}
}
\vspace*{0.7cm}
\centerline{
\includegraphics[scale=0.34,angle=-90]{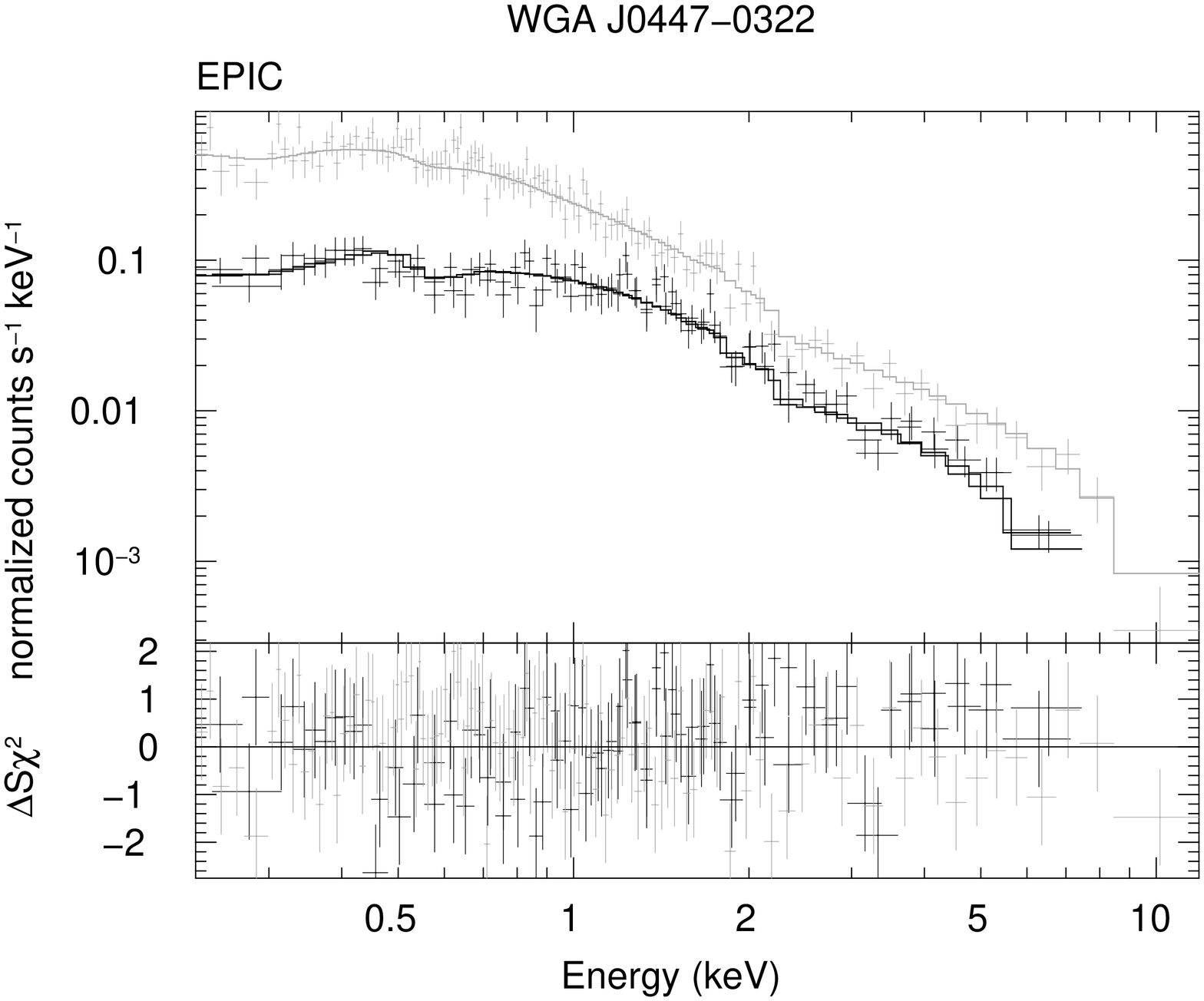}
\includegraphics[scale=0.34,angle=-90]{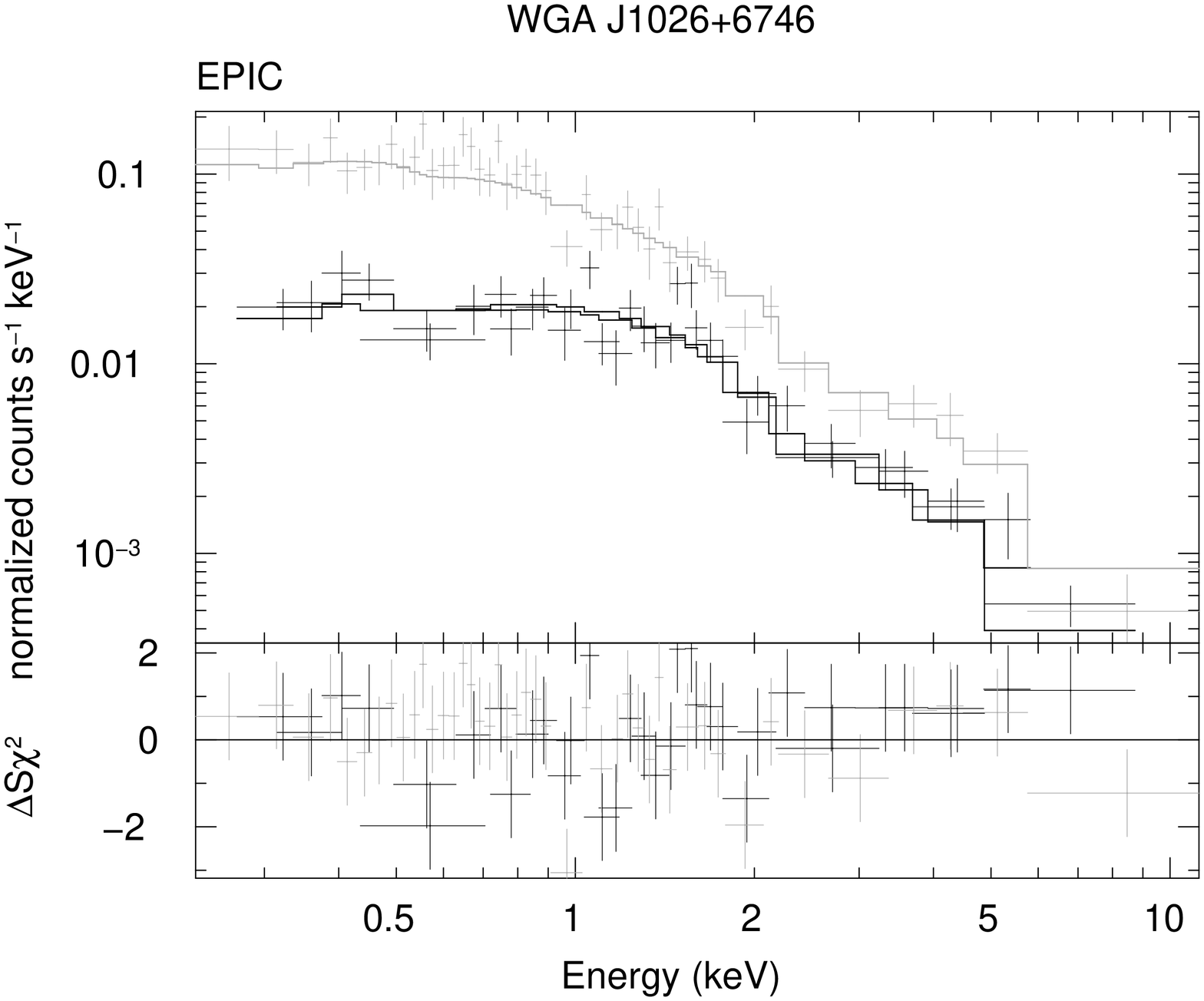}
}
\caption{\label{epicfit} EPIC MOS (black) and PN (grey) spectra
fit with a single power-law with Galactic hydrogen column density for
WGA J0110$-$1647, WGA J0304$+$0002, and WGA J1026$+$6746, and with a
broken power-law with Galactic hydrogen column density for WGA
J0447$-$0322.}
\end{figure*}

\begin{figure*}
\centerline{
\includegraphics[scale=0.32,angle=-90]{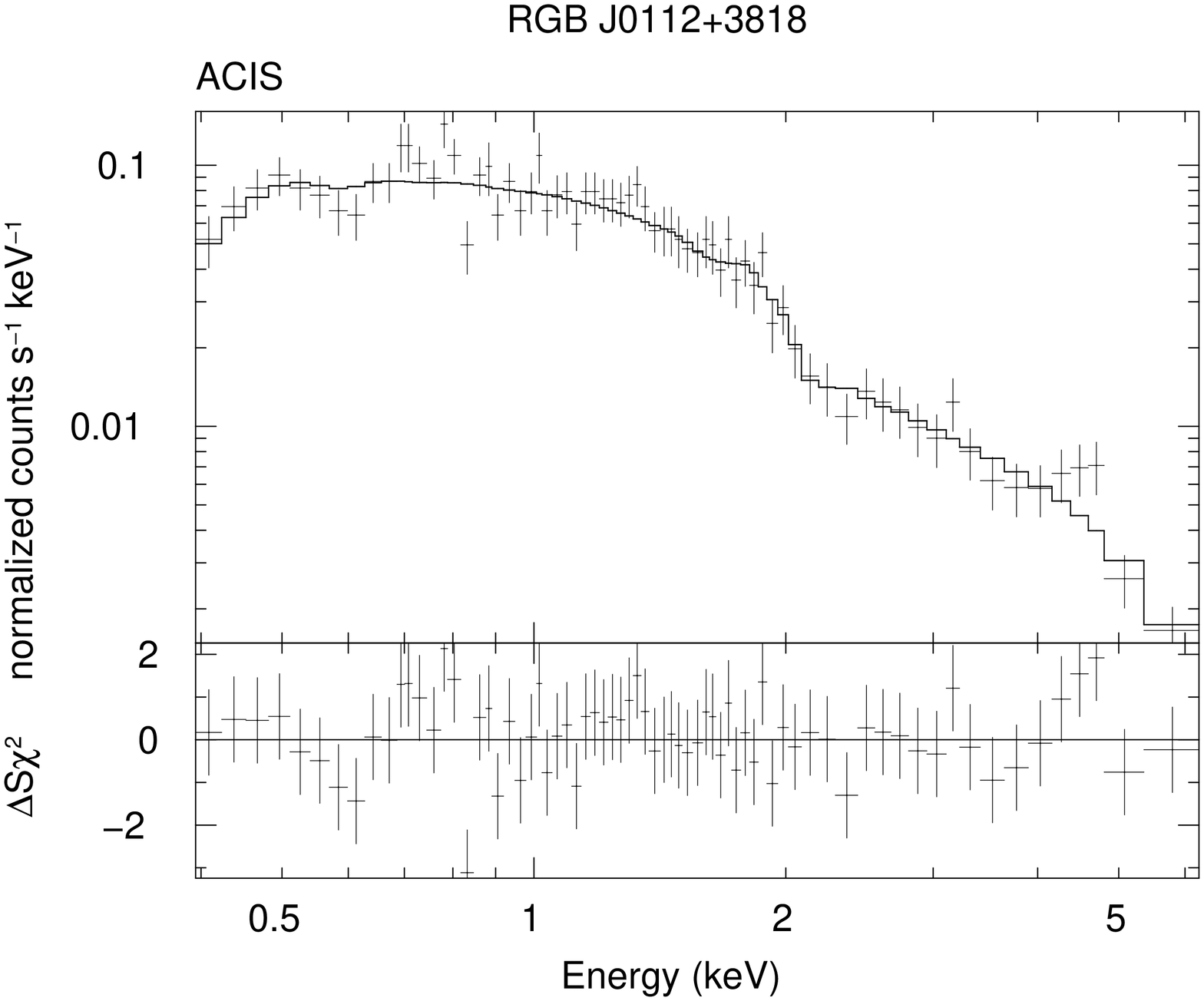}
\includegraphics[scale=0.32,angle=-90]{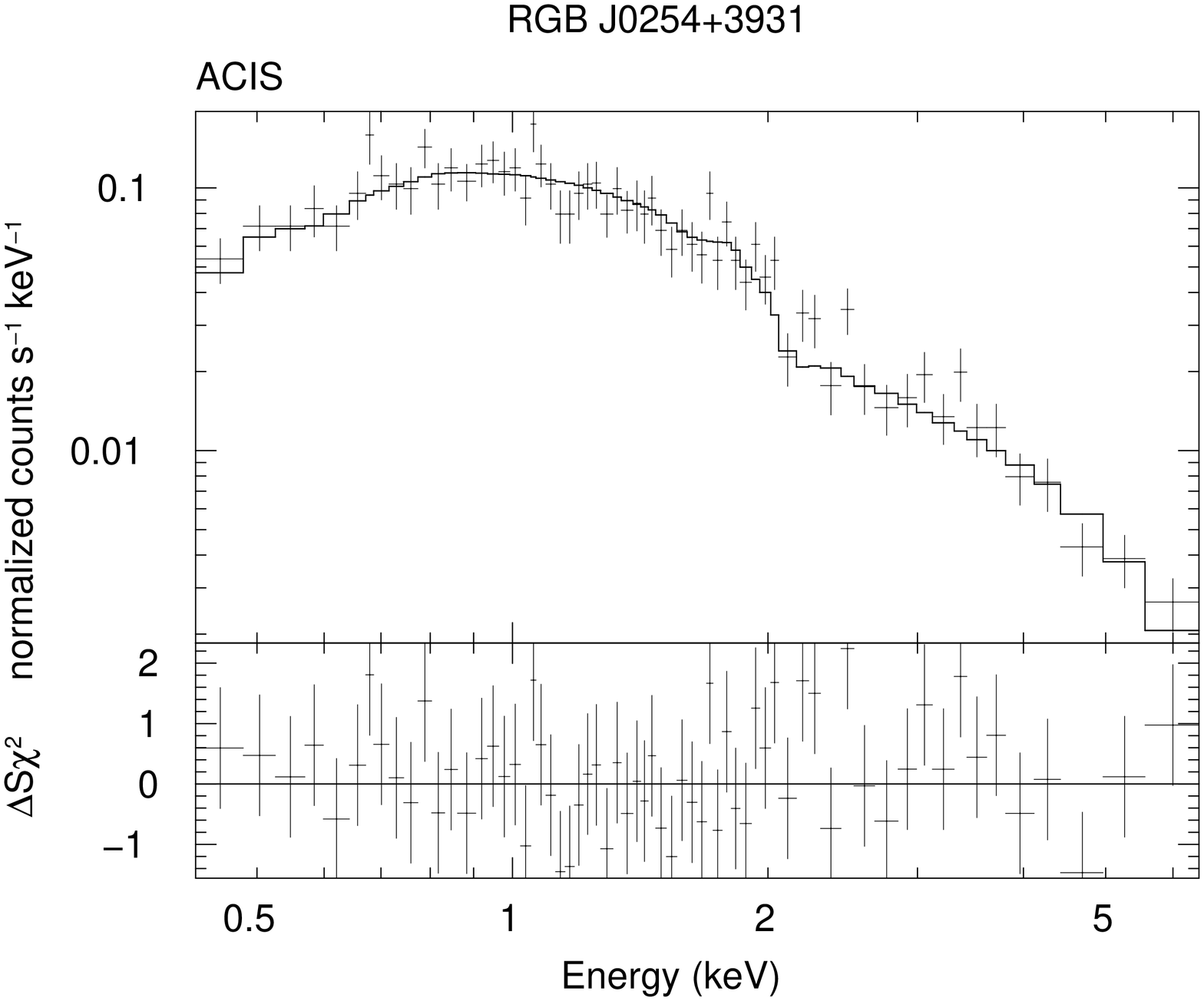}
}
\vspace*{0.6cm}
\centerline{
\includegraphics[scale=0.32,angle=-90]{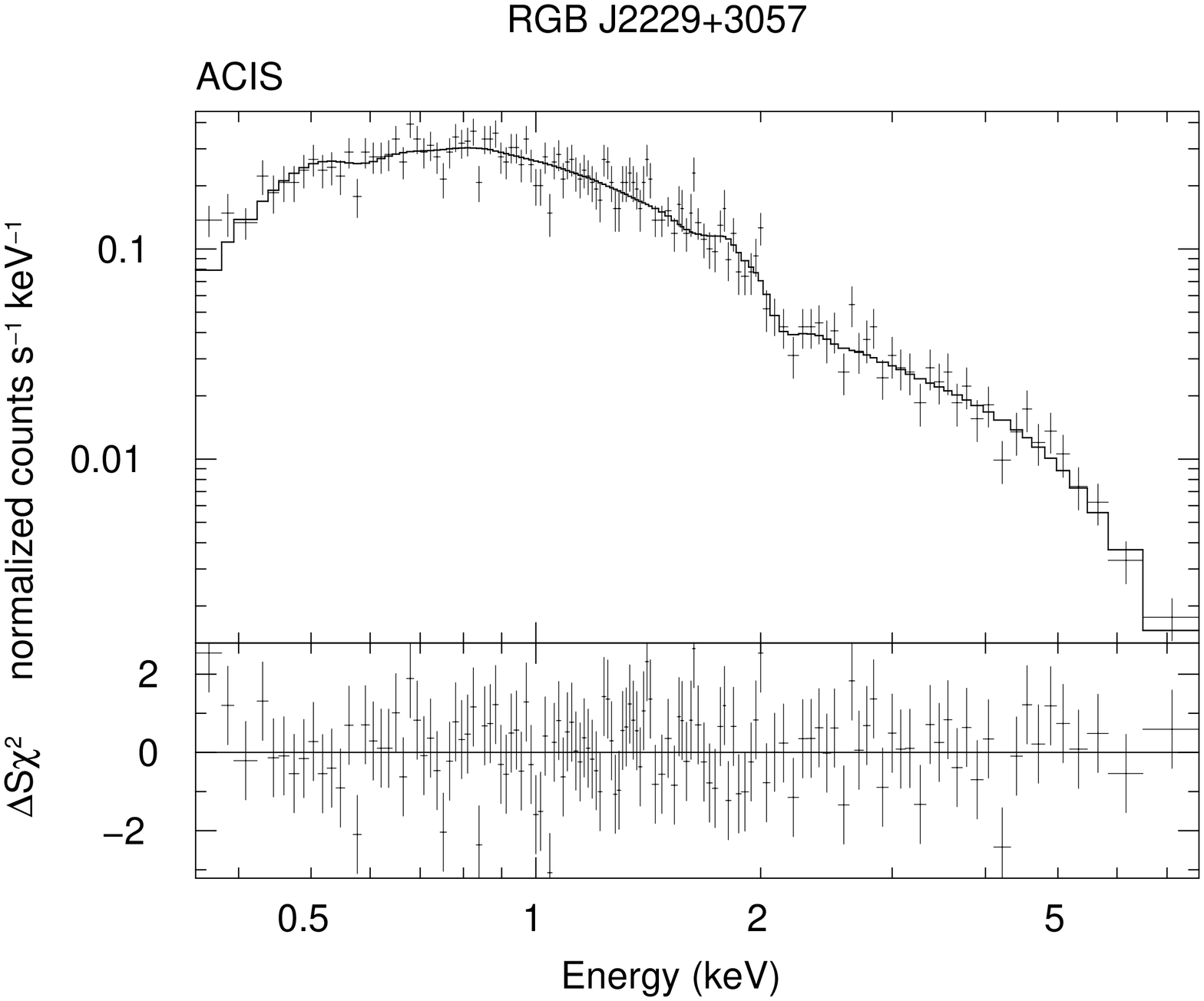}
\includegraphics[scale=0.32,angle=-90]{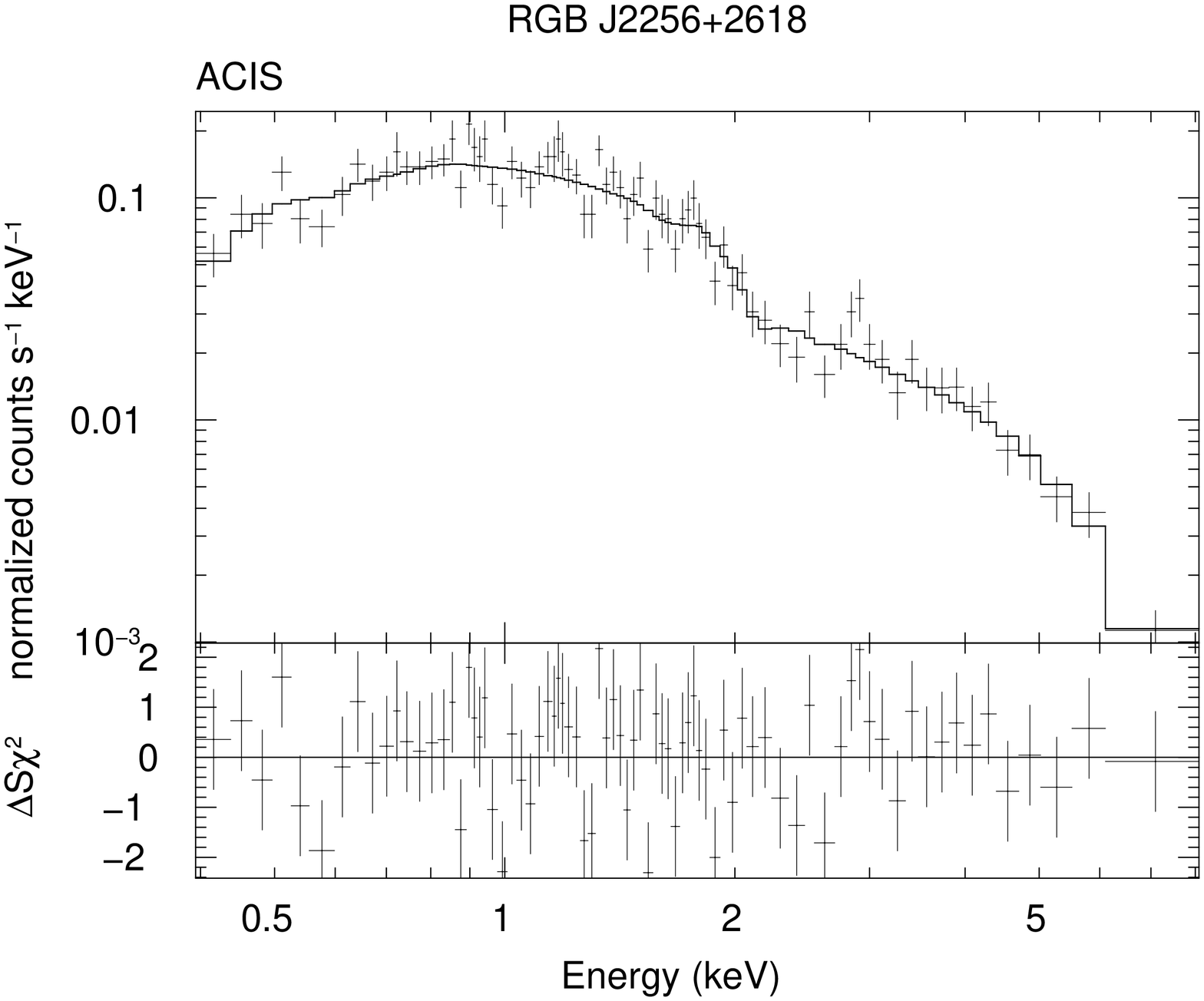}
}
\vspace*{0.6cm}
\centerline{
\includegraphics[scale=0.32,angle=-90]{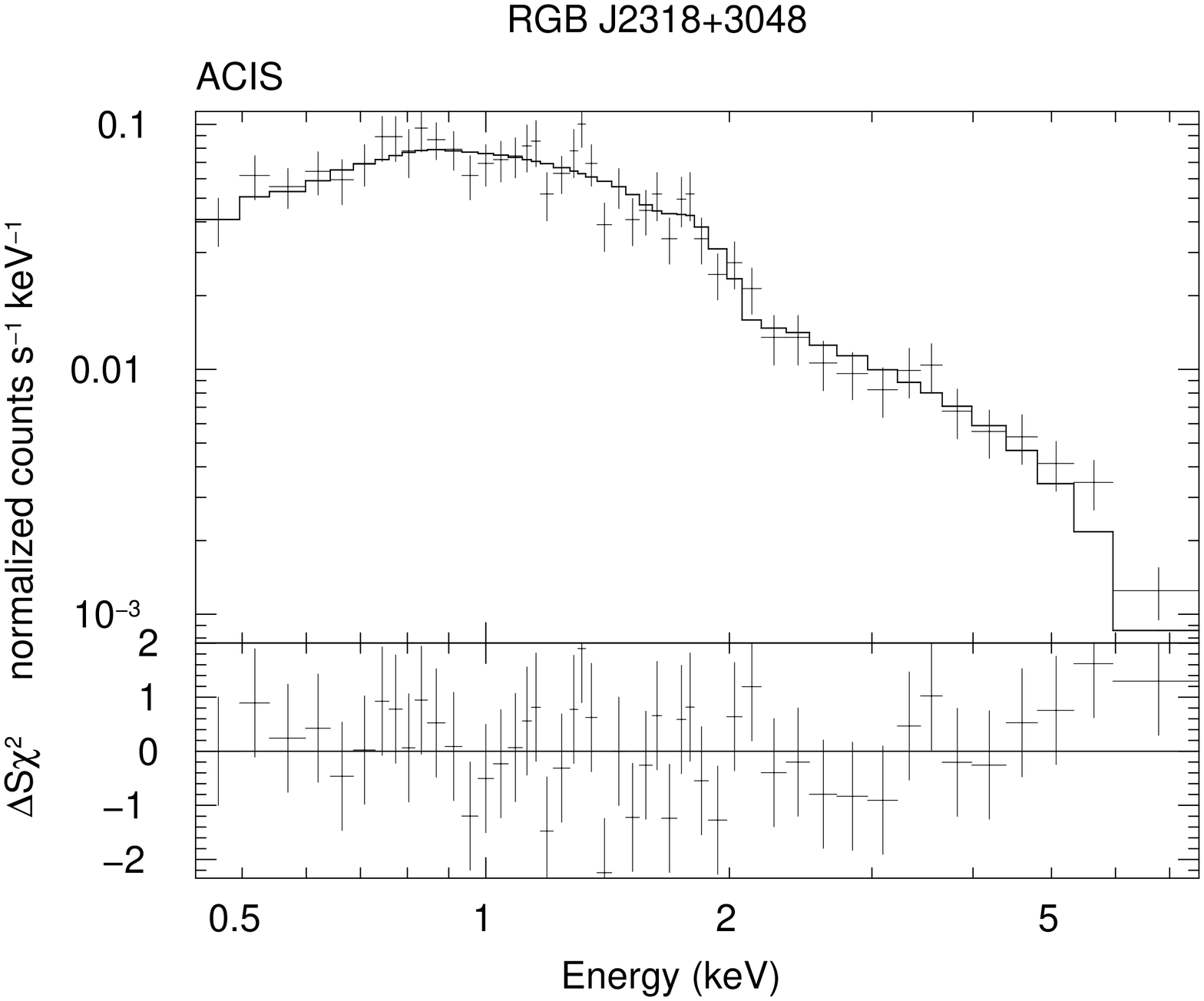}
\includegraphics[scale=0.32,angle=-90]{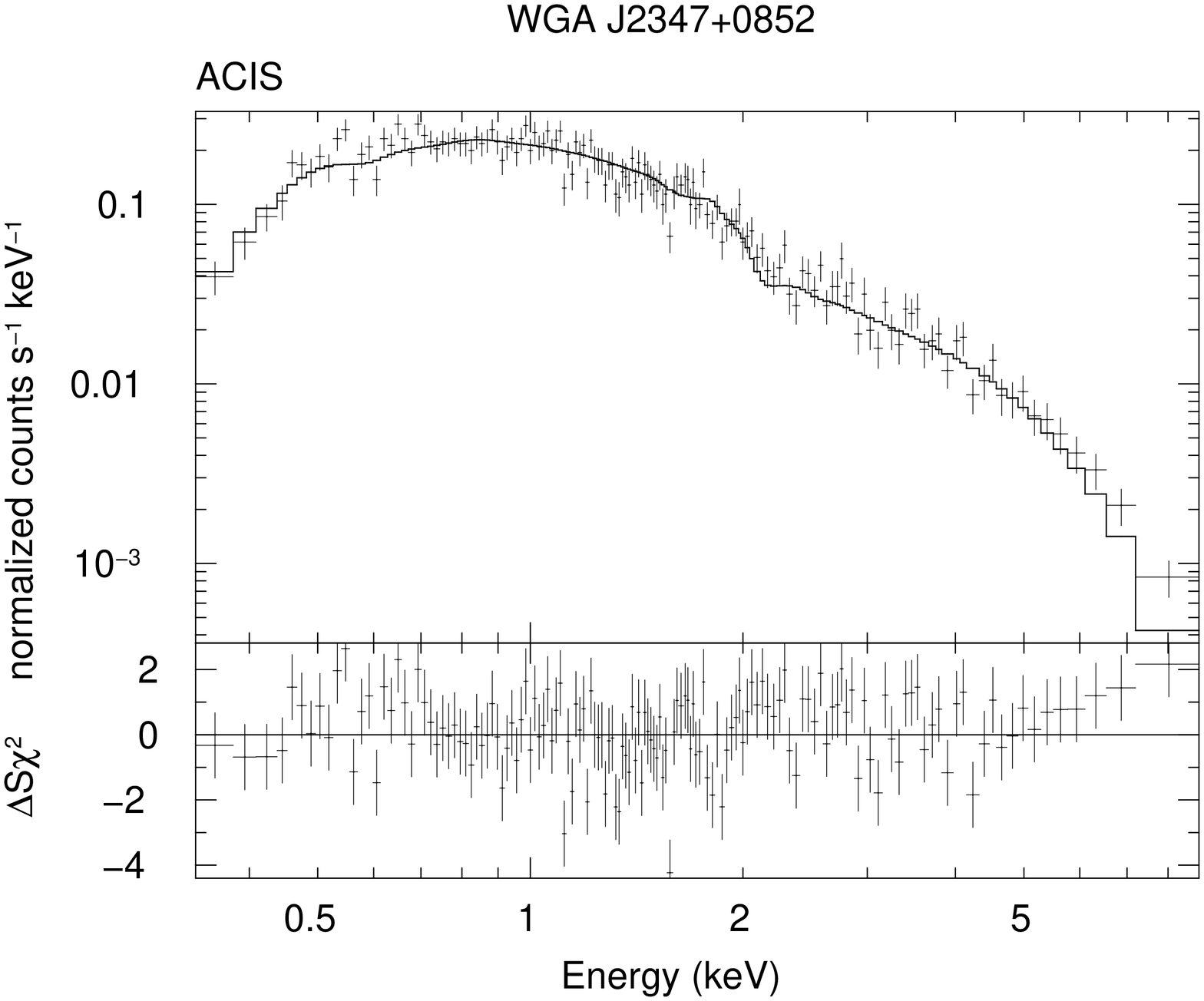}
}
\caption{\label{acisfit} ACIS spectra fit with a single
power-law with Galactic hydrogen column density for RGB J0254$+$3931,
RGB J2256$+$2618, RGB J2318$+$3048, and WGA J2347$+$0852, and with a
broken power-law with Galactic hydrogen column density for RGB
J0112$+$3818 and RGB J2229$+$3057.}
\end{figure*}

In the following we discuss for each object the results of the X-ray
spectral analysis. In Figs. \ref{epicfit} and \ref{acisfit} we show
for the {\sl XMM-Newton} and {\sl Chandra} observed sources,
respectively, the best-fit for the model which we consider the most
appropriate. This is a single power-law with Galactic absorption for
7/10 sources and a broken power-law with Galactic absorption for 3/10
sources, namely, WGA J0447$-$0322, RGB J0112$+$3818, and RGB
J2229$+$3057.

\vspace*{0.3cm}

{\sl WGA J0110$-$1647. $-$} The EPIC data are well fit by a single
power-law with Galactic absorption and a spectral index of
$\Gamma\sim2$. The data suggest that a broken power-law is a better
fit ($F$-test $>99.9\%$) than a single power-law. But since the break
is at relatively high energies ($E\sim7$ keV), the hard power-law
cannot be constrained by the present data. The soft spectral index is
in this case similar to the one obtained for the single power-law fit.
Additionally, there is a hint in the data of a soft excess; the fit is
marginally improved ($F$-test $\sim99\%$) assuming an absorption lower
than Galactic, but in this case the $N_{\rm H}$ is not well
constrained.

\vspace*{0.1cm}

{\sl WGA J0304$+$0002. $-$} The EPIC data are well fit by a single
power-law with Galactic absorption and a spectral index of
$\Gamma\sim1.9$. There is a hint in the data of a soft excess; the fit
is marginally improved ($F$-test $\sim98\%$) assuming an absorption
lower than Galactic. The fit is also marginally improved ($F$-test
$\sim99\%$) if we assume a broken power-law instead of a single
power-law. But in this case both the soft and hard spectral indices
are similar to the one obtained for the single power-law.

\vspace*{0.1cm}

{\sl WGA J0447$-$0322. $-$} The EPIC data are well fit by a broken
power-law with Galactic absorption. This fit represents a significant
improvement relative to a fit by a single power-law ($F$-test
$>99.9\%$). The resulting spectral indices of the soft and hard
power-laws are $\Gamma_{\rm soft}\sim2.3$ and $\Gamma_{\rm
hard}\sim1.7$, respectively, with an observed break at energy
$E\sim1.6$ keV.

\vspace*{0.1cm}

{\sl WGA J1026$+$6746. $-$} The EPIC data are well fit by a single
power-law with Galactic absorption and a spectral index
$\Gamma\sim1.8$.

\vspace*{0.1cm}

{\sl RGB J0112$+$3818. $-$} The ACIS data are well fit by a broken
power-law with Galactic absorption. This fit represents a significant
improvement relative to a fit by a single power-law ($F$-test
$>99.9\%$). The resulting spectral indices of the soft and hard
power-laws are $\Gamma_{\rm soft}\sim2.6$ and $\Gamma_{\rm
hard}\sim1.7$, respectively, with an observed break at energy
$E\sim0.9$ keV.

\vspace*{0.1cm}

{\sl RGB J0254$+$3931. $-$} The ACIS data are well fit by a single
power-law with Galactic absorption and a spectral index
$\Gamma\sim1.8$.

\vspace*{0.1cm}

{\sl RGB J2229$+$3057. $-$} The ACIS data are well fit by a broken
power-law with Galactic absorption. This fit represents a significant
improvement relative to a fit by a single power-law ($F$-test
$>99.9\%$). The resulting spectral indices of the soft and hard
power-laws are $\Gamma_{\rm soft}\sim2.3$ and $\Gamma_{\rm
hard}\sim1.7$, respectively, with an observed break at energy
$E\sim1.5$ keV.

\vspace*{0.1cm}

{\sl RGB J2256$+$2618. $-$} The ACIS data are well fit by a single
power-law with Galactic absorption and a spectral index
$\Gamma\sim1.7$.

\vspace*{0.1cm}

{\sl RGB J2318$+$3048. $-$} The ACIS data are well fit by a single
power-law with Galactic absorption and a spectral index of
$\Gamma\sim1.7$. There is a hint in the data of a soft excess; the fit
is marginally improved ($F$-test $\sim96\%$) assuming an absorption
lower than Galactic, but in this case the $N_{\rm H}$ is not well
constrained. The fit is also marginally improved ($F$-test $\sim99\%$)
if we assume a broken power-law instead of a single power-law. But in
this case both the soft and hard spectral indices are similar to the
one obtained for the single power-law.

\vspace*{0.1cm}

{\sl WGA J2347$+$0852. $-$} The ACIS data are well fit by a single
power-law with Galactic absorption and a spectral index of
$\Gamma\sim1.9$. There is a hint in the data of a soft excess; the fit
is marginally improved ($F$-test $\sim97\%$) assuming an absorption
lower than Galactic. The fit is also marginally improved ($F$-test
$\sim99\%$) if we assume a broken power-law instead of a single
power-law. But in this case the break is at relatively high energies
($E\sim5$ keV) and the resulting soft spectral index is similar to the
one obtained for the single power-law fit.

\section{The Spectral Energy Distributions} \label{sed}

In order to constrain the origin of the X-ray emission in our sources,
and in particular to understand if it is produced by the synchrotron
jet component, we have related it to the multiwavelength
SED. Additionally, we have estimated the flux contribution expected
from the two thermal emission components present in radio quasars,
namely, the accretion disk and the host galaxy.

\subsection{The Multiwavelength Data}

We have used the following uniform multiwavelength data, which, with
the exception of the magnitudes from the Optical Monitor (OM) on-board
{\sl XMM-Newton}, are non-simultaneous with our X-ray observations
(but, as noted below, in part simultaneous with each other):

\begin{enumerate}

\item
radio core fluxes at 1.4 GHz obtained with the VLA A array
\citep[from][reproduced in Table \ref{general}]{L06};

\item
near-IR $J$, $H$ and $Ks$ magnitudes (simultaneous to each other) from
the Two Micron All Sky Survey \citep[2MASS;][]{Skr06} Point Source and
Extended Source catalogs (see Table \ref{mag}, where the extended
fluxes are listed in parenthesis);

\item
optical red and blue magnitudes (simultaneous to each other) from the
Automatic Plate Measuring catalogs \citep[APM;][]{Irw94} (see Table
\ref{mag});

\item
optical magnitudes (simultaneous to each other) from the Sloan Digital
Sky Survey \citep[SDSS;][]{Ade06} Data Release 6 for WGA J0304$+$0002,
our only source included in this survey ($u=18.5$ mag, $g=18.1$ mag,
$r=18.2$ mag, $i=18.0$ mag, and $z=18.2$ mag);

\item
far- and near-UV magnitudes (simultaneous to each other) from the
Galaxy Evolution Explorer \citep[GALEX;][]{Mar05} Data Release 3
(see Table \ref{mag});

\item
optical and UV magnitudes (simultaneous to each other) from the OM for
the four sources observed with {\sl XMM-Newton} (see Table \ref{om});

\item
dereddened optical spectra from \citet{Per98} and \citet{L01} for the
DXRBS sources and from \citet{LM98} for the RGB sources; and

\item
unabsorbed {\sl ROSAT} X-ray flux densities at 1 keV (see Table
\ref{general}).

\end{enumerate}

\begin{figure*}
\centerline{
\includegraphics[scale=0.42]{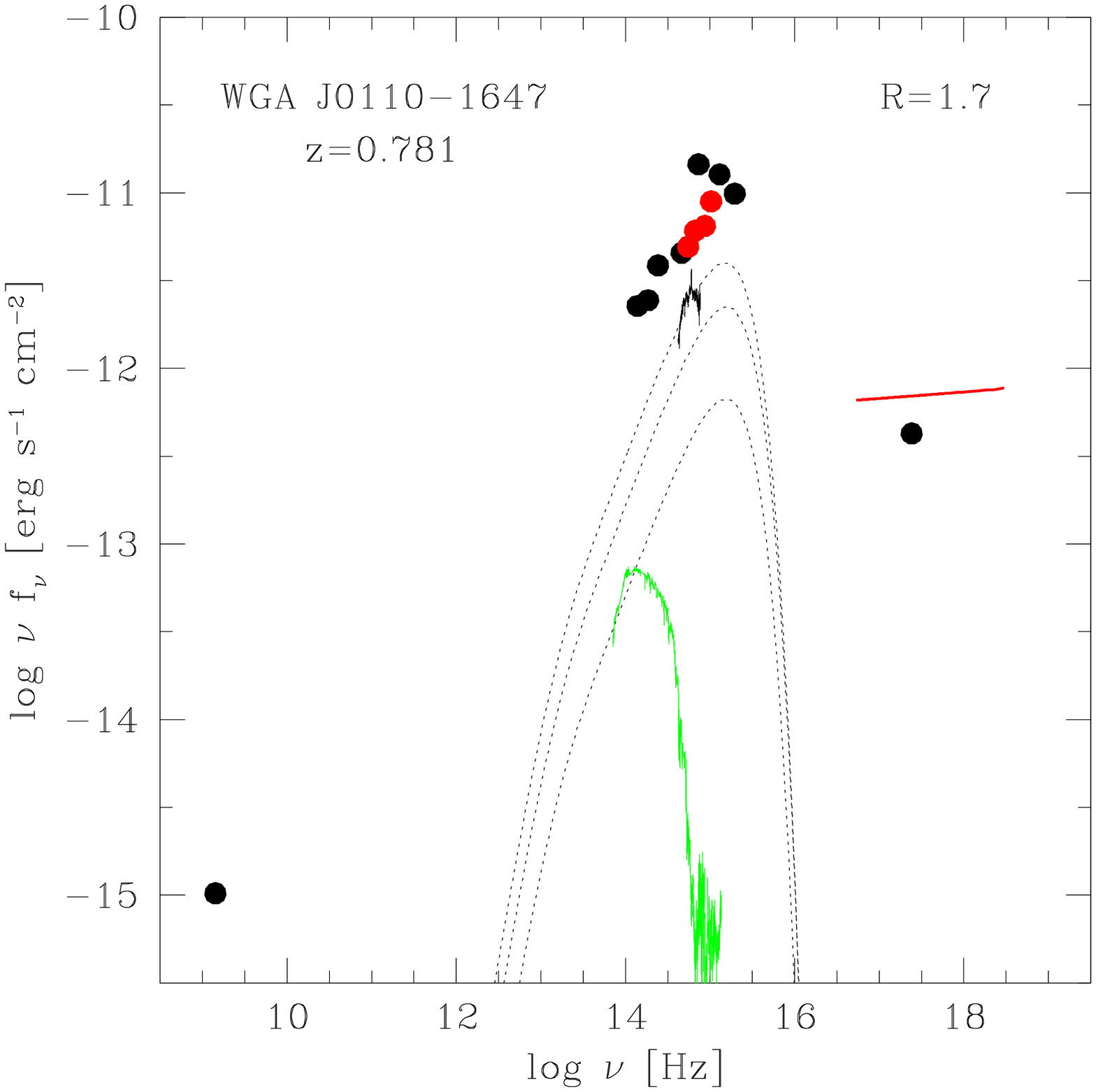}
\includegraphics[scale=0.42]{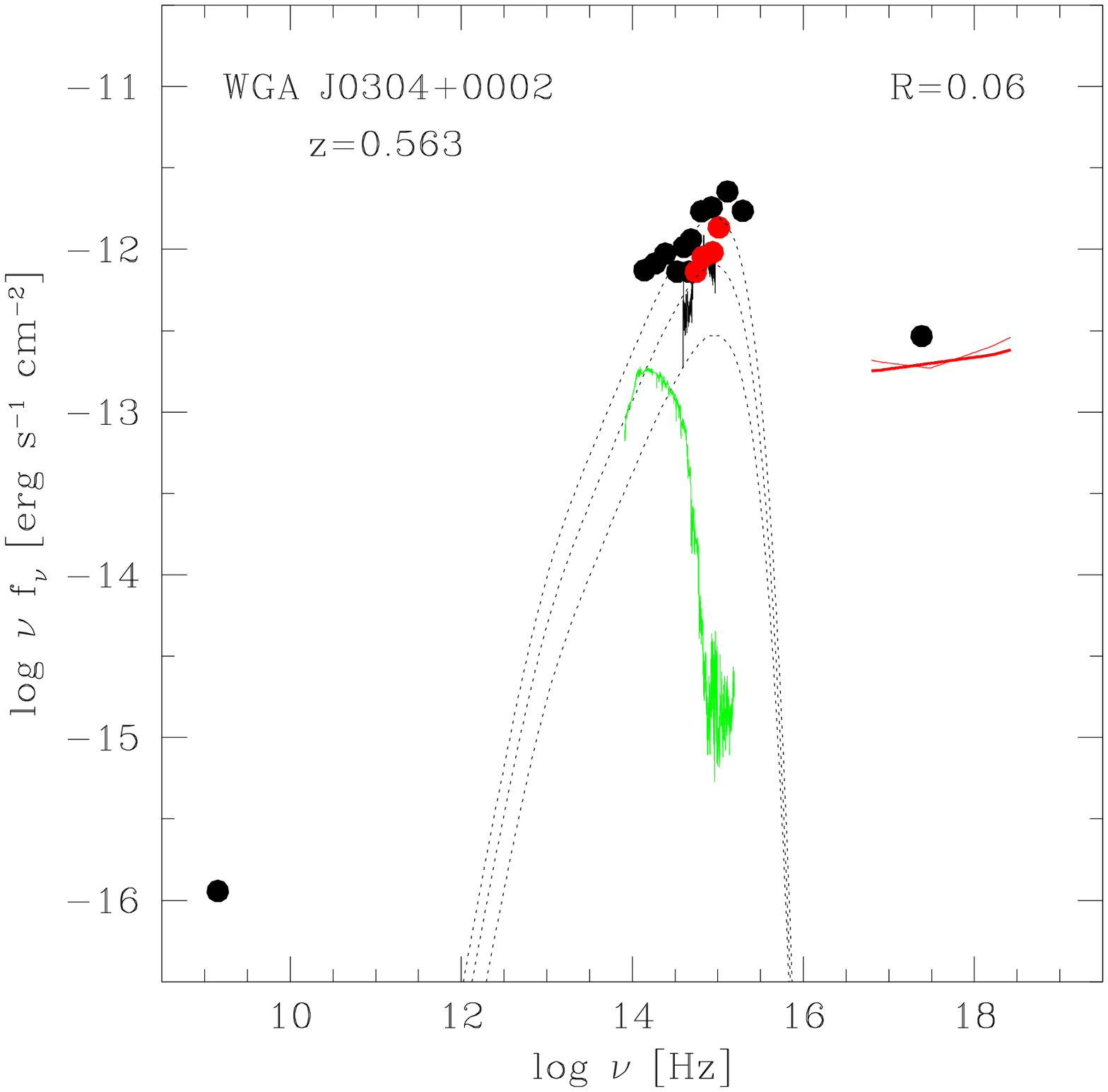}
}
\centerline{
\includegraphics[scale=0.42]{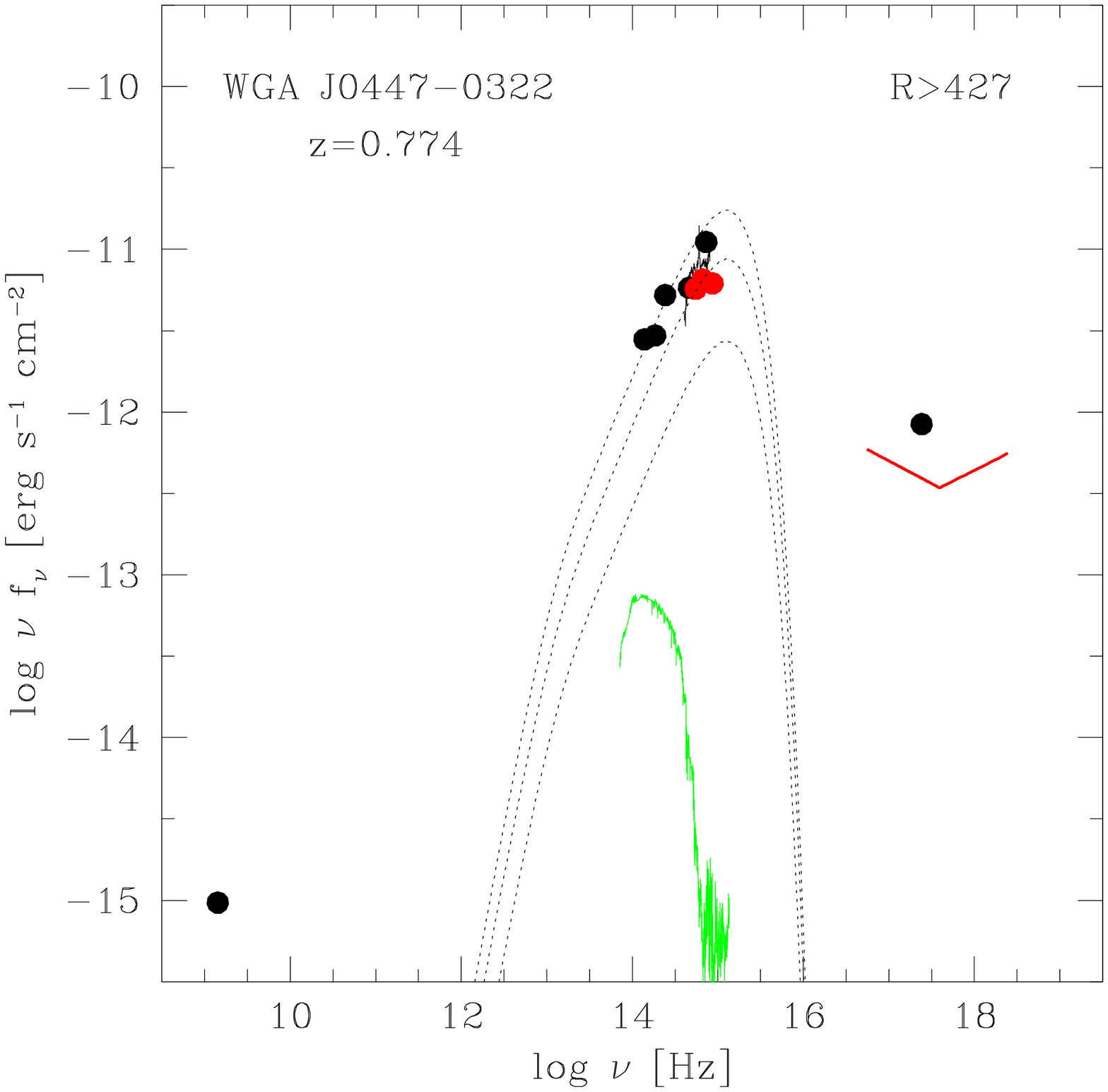}
\includegraphics[scale=0.42]{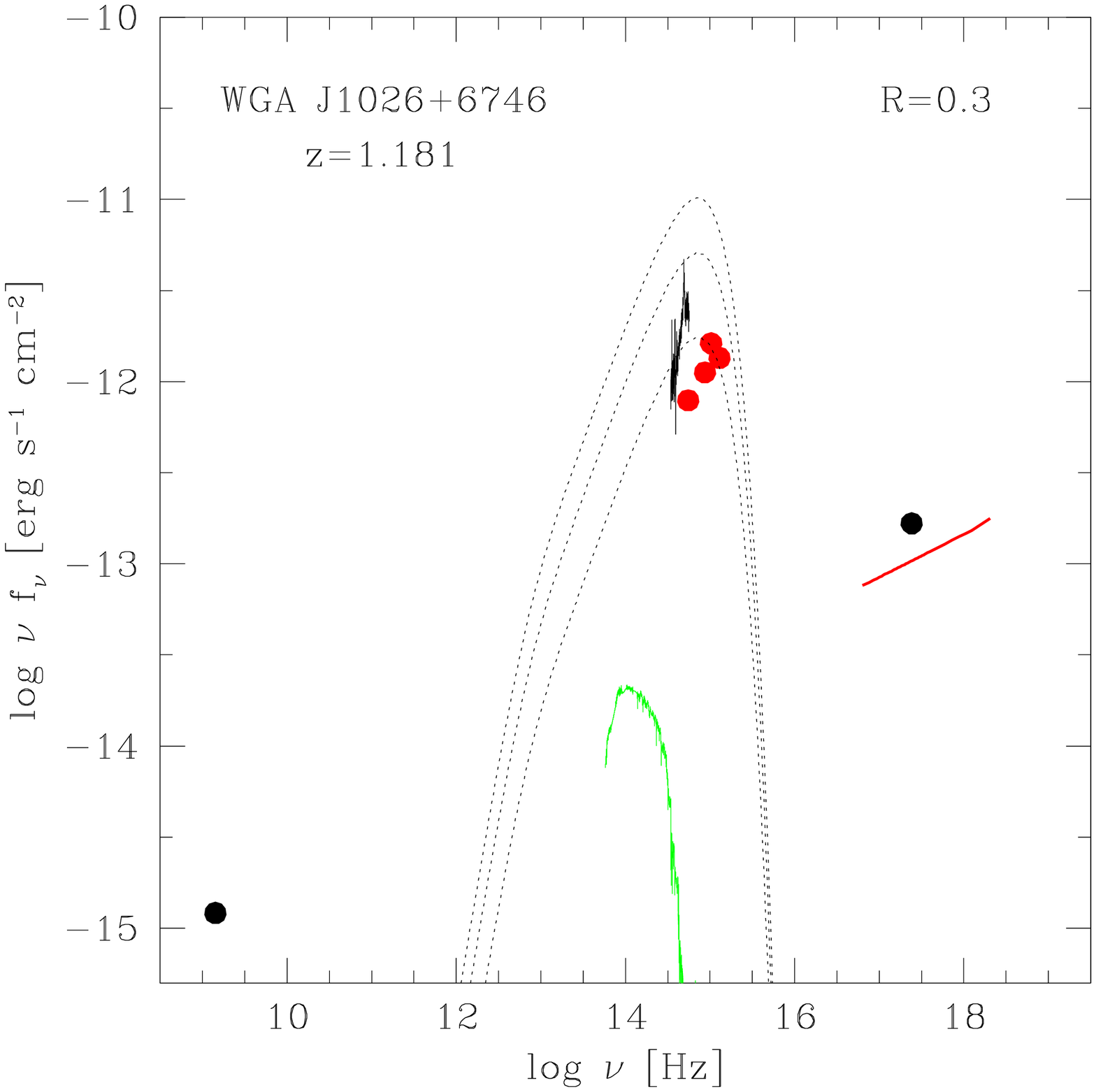}
}
\caption{\label{xmmsed} Observed multiwavelength SEDs of the
{\sl XMM-Newton} sources. The thick red solid line and circles
represent the X-ray spectrum and the magnitudes from the Optical
Monitor (OM), respectively. The broken power-law fits with marginal
significance are indicated by the thin red solid line. Black circles
represent non-simultaneous data at radio, near-IR, optical, UV and
X-ray frequencies. The extended near-IR fluxes are shown as green
circles. The black, solid line represents the optical spectrum. The
black, dotted curves indicate the accretion disk spectrum estimated
from the broad line region (BLR) luminosity and assuming a BLR
covering factor of $f_{\rm cov} = 5$, 10 and 30\% (from top to
bottom). The estimated host galaxy emission is represented by the
green, solid curve. See text for more details.}
\end{figure*}

The multiwavelength SEDs of our sources are shown in
Figs. \ref{xmmsed} and \ref{chandrased} for our {\sl XMM-Newton} and
{\sl Chandra} observed sources, respectively. We have plotted the SEDs
in the observer's frame as log $\nu f_{\nu}$ vs. log $\nu$. In such a
representation the peak indicates directly at which frequency most of
the energy is emitted. All magnitudes have been corrected for Galactic
extinction prior to their inclusion in Figs. \ref{xmmsed} and
\ref{chandrased}. For this purpose we used the analytical expression
for the interstellar extinction curve of \citet{Car89} and a parameter
of $R_{\rm V} = 3.1$ to transform the $A_{\rm V}$ values to
$A_{\lambda}$ values. We note that the optical spectra were obtained
through a relatively small ($\sim 1''$) aperture, and, therefore, in
particular for nearby objects, are not expected to account for all the
flux from extended emission components, such as, e.g., the host galaxy
(see Section \ref{host}).

\subsection{The Thermal Emission Components} \label{thermal}

In addition to the relativistically beamed, non-thermal jet emission,
two thermal emission components are expected to be present in radio
quasars: the accretion disk and the host galaxy. We have estimated
their contribution to the multiwavelength SED as follows.

\subsubsection{The Accretion Disk}

We have calculated accretion disk spectra assuming a steady
geometrically thin, optically thick accretion disk. In this case the
emitted flux is independent of viscosity and each element of the disk
face radiates roughly as a blackbody with a characteristic
temperature, which depends only on the mass of the black hole, $M_{\rm
BH}$, the accretion rate, $\dot{M}$, and the radius of the innermost
stable orbit \citep[e.g.,][]{Peterson, FKR}. We have adopted the
Schwarzschild geometry (non-rotating black hole) and for this the
innermost stable orbit is at $r_{\rm in} = 6 \cdot r_{\rm g}$, where
$r_{\rm g}$ is the gravitational radius defined as $r_{\rm g} = G
M_{\rm BH}/c^2$ with $G$ the gravitational constant and $c$ the speed
of light. Furthermore, we have assumed that the disk is viewed
face-on.

The accretion disk spectrum is fully constrained by the two
quantities, accretion rate and mass of the black hole, which we have
estimated as follows. We have calculated the accretion rate from the
luminosity of the broad emission lines using the two relations: (1)
$L_{\rm ion} = \epsilon \dot{M} c^2$, where $L_{\rm ion}$ and
$\epsilon$ are the total ionizing power of the disk and the efficiency
for converting matter to energy, respectively, with $\epsilon \sim$6\%
in the case of a Schwarzschild black hole; and (2) $L_{\rm ion} =
f_{\rm cov}^{-1} L_{\rm BLR}$, where $f_{\rm cov}$ and $L_{\rm BLR}$
are the BLR covering factor and luminosity, respectively. We have
calculated BLR luminosities from the fluxes of the observed broad
emission lines as described in \citet{Cel97}. The optical spectra of
our sources cover mostly the broad emission lines \MgII~$\lambda 2798$
and H$\beta$, and in two cases also H$\alpha$. The BLR covering factor
is not well known, but is derived to be in the range of $\sim 5-30$\%
\citep[][and references therein]{Mai01} and in general a canonical
value of $\sim 10$\% is assumed \citep[][]{Peterson}. 

Assuming that the BLR is gravitationally bound, the mass of the black
hole can be estimated based on the virial theorem from the width of a
broad emission line and the ionizing luminosity. In this case the
ionizing power is used as a surrogate for the BLR radius. Following
\citet{Delia03} we have calculated black hole masses for 6/10 sources
using the full width at half maximum (FWHM) of H$\beta$ and the
ionizing power $L_{\rm ion}$, the latter calculated from the BLR
luminosity as above. For 4/10 sources, for which the optical spectrum
does not cover the wavelength of H$\beta$, we have used the \MgII~FWHM
as a substitute for that of H$\beta$. As \citet{McL02} have shown, the
widths of H$\beta$ and \MgII~are almost identical, indicating that, as
expected from their (similar) ionization potentials, these two broad
emission lines are produced at similar radii.

In Table \ref{acctable} we list for our sources the observed BLR
luminosities and line widths along with the accretion rates and black
hole masses calculated assuming $f_{\rm cov}=10$\%. We note that in
two sources, namely, WGA J0304$+$0002 and WGA J2347$+$0852, H$\beta$
has a clear narrow component, which we have subtracted before
measuring the width. In Figs. \ref{xmmsed} and \ref{chandrased} we
show accretion disk spectra (black, dotted curves) for three sets of
accretion rates and black holes masses corresponding to $f_{\rm
cov}=5, 10$ and 30\% (from top to bottom).

\subsubsection{The Host Galaxy} \label{host}

The host galaxies of radio quasars are bright ellipticals that exhibit
a relatively narrow range in luminosity \citep[standard deviation
$\sigma \sim 0.1-0.6$ mag; e.g.,][]{McL04a, Dun03, Dun93,
Kot98}. Therefore, we have estimated the contribution from this
emission component using the elliptical galaxy template of
\citet{Man01}, which extends from near-IR to UV frequencies, and
assuming a fixed absolute luminosity in the (rest-frame) $R$-band of
$M_R=-23.2$ mag. This is the average value obtained by \citet{McL04a}
for the host galaxy luminosities of a sample of 41 radio galaxies
imaged with the {\sl Hubble Space Telescope} in the (observer's frame)
{\sl I}-band. The green, solid curves in Figs. \ref{xmmsed} and
\ref{chandrased} show our results.

\section{Results and Discussion}

We now address the main question of this paper: Have we found among
this subsample of ``X-ray loud'' radio quasars as expected
strong-lined blazars with $\nu_{\rm peak} > 10^{15}$ Hz and thus
X-rays dominated by synchrotron emission?

\subsection{Is the X-ray Emission Synchrotron?}

In a representation of the SED as a logarithmic $\nu f_{\nu} - \nu$
plot (as in Figs. \ref{xmmsed} and \ref{chandrased}) the two main jet
components, synchrotron and inverse Compton radiation, form two humps
peaking at lower and higher energies, respectively. The synchrotron
emission in our sources presumably peaks at UV/soft-X-ray
frequencies. Therefore, if the X-ray emission was synchrotron in
origin, it would be necessarily the part of the synchrotron spectrum
after the emission peak and thus curved downward. In the simplest
approximation this means that in this case a single power-law with
slope $\Gamma > 2$ is expected.

In our sample, 7/10 sources are well fit by a single power-law,
however, all with values $\Gamma < 2$. Therefore, synchrotron X-ray
emission is not observed in the large majority of our sources. Three
sources, namely, WGA J0447$-$0322, RGB J0112$+$3818, and RGB
J2229$+$3057, are well fit by a broken power-law. The best-fits give
in all three cases soft and hard spectral indices $\Gamma_{\rm
soft}>2$ and $\Gamma_{\rm hard}<2$, respectively. This suggests that,
as in ``intermediate'' BL Lacs, we observe the spectral transition
between the synchrotron and inverse Compton jet components, which in
these radio quasars occurs at relatively low X-ray energies of $E \sim
1 - 1.5$ keV.

\subsection{Is the Synchrotron Peak in the UV?} \label{UVpeak}

In the seven sources well fit by a single power-law synchrotron
emission does not dominate at X-ray frequencies. Therefore, the X-ray
emission is either produced by the inverse Compton jet component as in
``classical'' FSRQ, indicating a relatively low $\nu_{\rm peak}$, or
by an emission component unrelated to the jet, indicating that the jet
is generally weak. The first scenario is excluded by the $\alpha_{\rm
rx}$ values of our sources, which are much lower than the typical
value of $\alpha_{\rm rx} \sim 0.85$ of ``classical'' FSRQ
\citep{P03}. The latter scenario seems likely. As Table \ref{general}
shows, these FSRQ are either lobe-dominated (i.e., have $\log R=\log
L_{\rm core}/L_{\rm ext}<0$, where $L_{\rm core}$ and $L_{\rm ext}$
are the radio core and extended luminosity, respectively; 3/7 objects)
or are only slightly core-dominated (i.e., have $0<\log R<0.3$; 4/7
objects), which suggests that their jets are weakly beamed. In this
case a $\nu_{\rm peak} \ga 10^{15}$ Hz cannot be excluded, but then
these sources were not selected as high-energy peaked blazar
candidates based on their jet SED.

Out of the three sources well fit by a broken power-law, only WGA
J0447$-$0322 has an SED sampled simultaneously at optical/UV and X-ray
frequencies (by {\sl XMM-Newton}) and an extrapolation of its soft
X-ray power-law to lower energies predicts fluxes a factor of $\sim 3$
below the observed OM magnitudes. This discrepancy suggests that a
single emission component cannot account for the fluxes at both
frequencies. Therefore, either the soft-X-ray flux is synchrotron and
the optical/UV fluxes are not, in which case the accretion disk could
produce the optical/UV emission (see Section \ref{selection}), or the
optical/UV magnitudes sample the synchrotron jet component (which has
$\nu_{\rm peak} \ga 10^{15}$ Hz), but the observed soft X-ray spectral
slope is too flat (we refer here and in the following to ``flat'' and
``steep'' in $\log \nu f_{\nu} - \log \nu$) to be synchrotron emission
alone.

A similar argument for an additional emission component can also be
made for RGB J2229$+$3057, but less stringently so, since its
optical/UV magnitudes and X-ray spectrum are not
simultaneous. Nevertheless, it is likely that RGB J2229$+$3057 and WGA
J0447$-$0322 are similar cases, since their best-fit X-ray broken
power-laws have similarly flat spectral slopes and low break
energies. For RGB J0112$+$3818, an extrapolation of the soft X-ray
power-law to lower energies predicts fluxes similar to the observed
optical magnitudes. However, also for this source the optical and
X-ray observations are not simultaneous and, moreover, its soft X-ray
power-law is not as well constrained as that of WGA J0447$-$0322 and
RGB J2229$+$3057.

The fact that an emission component in addition to synchrotron is
required at either optical/UV or soft X-ray frequencies (or both) for
WGA J0447$-$0322 and RGB J2229$+$3057 means that these sources are not
strong-lined analogs to HBL. On the other hand, if instead the analogy
to ``intermediate'' BL Lacs holds for the three sources well fit by a
broken power-law, as suggested by their X-ray spectrum, we expect
their synchrotron emission peak to be at relatively low frequencies
\citep[typically $\nu_{\rm peak} \sim 10^{14}$ Hz; e.g.,][]{Gio02}.

\subsection{Revisiting the Selection Criteria} \label{selection}

Given that in none of our sources we detected an X-ray spectrum
dominated by jet synchrotron emission the important question arises:
Why do the selected radio quasars have multiwavelength flux ratios
typical of HBL? Our method selected sources based on their high X-ray
to radio flux ratios (low $\alpha_{\rm rx}$) as well as high optical
to radio flux ratios (low $\alpha_{\rm ro}$). Therefore, if it is not
a jet synchrotron spectrum peaking at UV/soft-X-ray frequencies and
dominating the X-rays that increases both the optical and X-ray fluxes
relative to that at radio frequencies (as in HBL), a different
emission component must cause a similar effect.

Since our sources are radio quasars with strong broad emission lines,
the most likely component increasing the optical flux relative to that
at radio frequencies is the accretion disk. However, low-redshift
sources have usually low-luminosity AGN and their optical flux could
be instead enhanced by the host galaxy emission. Our estimate of the
fluxes of these two thermal emission components (Section
\ref{thermal}) suggests that the accretion disk can contribute
significantly to the optical fluxes of the majority of our sources
(8/10 objects; see Figs. \ref{xmmsed} and \ref{chandrased}). The host
galaxy, on the other hand, appears to dominate the optical magnitudes
of only the two most nearby objects, namely, RGB J2256$+$2618
($z=0.121$) and RGB J2318$+$3048 ($z=0.103$).

The estimated accretion disk spectrum approximate well the optical
spectra of 5/8, 2/8 and 1/8 objects if a BLR covering factor of
$f_{\rm cov}=5$, 10 and 30\%, respectively, is assumed. In 5/8 sources
we observe moderate variability (by factors of $\sim 2-5$) between the
optical spectra and the magnitudes. This variability behavior is
compatible with that of radio-quiet quasars \citep[e.g.,][]{Pal94,
Giv99, Van04, Wil05} and, as in these, it could be due to a change in
accretion rate \citep[][and references therein]{Pere06}. In this
respect we note that in particular the simultaneous OM magnitudes (and
in the case of WGA J0304$+$0002 also the simultaneous SDSS magnitudes)
mimic the slope of the estimated accretion disk spectrum.

Our interpretation that the host galaxy dominates the optical fluxes
of the two most nearby objects is supported by the fact that they
appear extended on digitized images from the Second Palomar Sky Survey
(POSS2). Additionally, their extended near-IR magnitudes lie well
above those from the 2MASS Point Source catalog and comply with the
estimated flux and spectral shape of the host galaxy. Similarly, their
optical spectra mimic the expected spectral shape of the host galaxy,
but the small aperture used in spectroscopy underestimates its flux.

Generally, accretion disk emission is expected to be unimportant in
FSRQ. Their jets are assumed to be strongly beamed and to dominate the
emission even at UV frequencies, where the accretion disk spectrum
peaks \citep{Delia03}. However, \citet[][see their Fig. 11]{L06} have
recently shown that a flat radio spectrum does not always ensure that
the radio source is highly core-dominated and thus strongly
beamed. This means that in some (few) FSRQ the accretion disk emission
could dominate over the (weakly beamed) jet spectrum. As already
discussed in Section \ref{UVpeak}, our sample seems to fall in this
category. In this respect we note that the {\it Chandra} sources were
chosen also because of their extended radio emission. Nevertheless,
among the {\it XMM-Newton} sources, which were chosen based on their
estimated high $\nu_{\rm peak}$, 2/4 sources are lobe-dominated and
one source has log $R<0.3$.

Given that thermal accretion disk emission is pronounced in our
sources one can assume that their X-ray fluxes are increased relative
to that at radio frequencies by the same component that renders also
radio-quiet quasars strong X-ray emitters. In support of the notion
that both the optical/UV and X-ray fluxes in our FSRQ have a similar
(non-jet) origin as in radio-quiet quasars are also their $\alpha_{\rm
ox}$ values (see Fig. \ref{aroaox}), which are not inconsistent with
those of radio-quiet quasars \citep[e.g.,][]{Gio99a, Str05}.

The X-ray emission of radio-quiet quasars is usually dominated by a
power-law with a flat spectral index [e.g., $\Gamma = 1.89 \pm 0.11$
\citep{Pic05}; $\Gamma = 1.92^{+0.09}_{-0.08}$ \citep{Just07}], which
is generally interpreted as the Comptonized hot corona of the
accretion disk. The seven sources well fit by a single power-law have
an average X-ray spectral index of $\Gamma = 1.81 \pm 0.04$, similar
($\la1 \sigma$ level) to the averages found for radio-quiet
quasars. But more than half of these sources (4/7 objects) show also a
hint of a broken power-law in their X-ray spectra (thin, red solid
lines in Figs. \ref{xmmsed} and \ref{chandrased}; see also Section
\ref{discuss}). In particular, in WGA J0110$-$1647, WGA J2347$+$0852
and RGB J2318$+$3048 a different (steep) component seems to emerge at
higher X-ray energies ($E \sim 7$, 5 and 3 keV, respectively), and in
WGA J0304$+$0002 a broken power-law similar in break energy to that
observed in WGA J0447$-$0322 and RGB J2229$+$3057 but with much
flatter both soft and hard X-ray spectral indices is evident at a
lower significance level.

Especially the case of WGA J0304$+$0002 suggests that in the selected
FSRQ the flat X-ray power-law component present in radio-quiet quasars
adds onto the jet SED, leading to an increase in total X-ray flux and
to a flattening of both soft and hard X-ray spectral slopes. Then in
sources where this component is particularly strong relative to the
jet, such as, e.g., WGA J0110$-$1647, the jet inverse Compton
component can emerge only at higher energies.

To a first approximation we expect in this scenario that the stronger
the relativistic beaming, the stronger the jet contribution and,
therefore, the less flattened the X-ray spectral slopes. And indeed,
two of the three sources well fit by a broken power-law are also those
with the highest radio core-dominance values in the sample (log
$R\ga2.6$ for WGA J0447$-$0322 and log $R \sim 0.7$ for RGB
J0112$+$3818).

The additional flat X-ray power-law will dominate the jet SED in
particular around the spectral transition point between the
synchrotron and inverse Compton components. This is exemplified by our
simulations in Fig. \ref{corona} for which we have used the average
blazar SED of \citet{Gio05} and to which we have added a component
with a spectral index of $\Gamma=2$, assuming flux ratios at the jet
SED transition point of 1, 5, 10, 20, 30, 40, 50, 100, and
150. Therefore, an increased X-ray flux relative to that at radio
frequencies is expected to be more pronounced in FSRQ whose spectral
break is at or around the selection X-ray energy. In this respect we
note that 4/10 sources in our sample and 2/4 sources observed by
\citet{P02} have a detected jet SED transition point around $E=1$ keV
(the X-ray energy used to calculate the $\alpha_{\rm rx}$ and
$\alpha_{\rm ox}$ values).

\subsection{A Refined Selection Technique}

\citet{P02} and this work presented X-ray spectroscopy for a total of
14 ``X-ray loud'' radio quasars and only one source, namely, RGB
J1629$+$4008, had X-rays dominated by synchrotron radiation. We have
shown that this modest success in finding radio quasars with
synchrotron X-rays is due to the adopted selection criteria, which
efficiently select BL Lacs with synchrotron X-rays, but do not work
well for radio quasars. Nevertheless, the important question remains:
does a large population of radio quasars with synchrotron X-rays
exist, and, if yes, how can they be efficiently found?

An efficient selection of radio quasars candidate to have synchrotron
X-rays could be based on the diagram presented by \citet[][see their
Fig. 12]{L06}, which plots log $L_{\rm core}/L_{\rm x}$ versus log
$R$, where $L_{\rm core}$ and $L_{\rm x}$ are the radio core and total
X-ray luminosity, respectively. In this plot high-energy peaked FSRQ
are expected to separate from the rest of the population, since they
should be those strongly core-dominated radio quasars (i.e., with log
$R\gg0$) that have the lowest $L_{\rm core}/L_{\rm x}$ ratios.

For BL Lacs values of log $L_{\rm core}/L_{\rm x} \la 6$ are known to
be indicative of synchrotron X-rays \citep{P96}. However, based on the
results in the previous section lower values must be chosen for radio
quasars. E.g., whereas RGB J1629$+$4008, which is strongly
core-dominated (log $R>1.5$), has a ratio of log $L_{\rm core}/L_{\rm
x} \sim 4.5$ \citep{L06}, the two strongly core-dominated sources in
our sample, namely, WGA J0447$-$0322 and RGB J0112$+$3818, have ratios
of log $L_{\rm core}/L_{\rm x} \sim 5$ and $\sim 5.5$ \citep{L06},
respectively, which are apparently not low enough to be indicative of
synchrotron X-rays.

But it remains to be shown if radio quasars with synchrotron X-rays
exist in large numbers and both a positive and a negative result will
reveal highly significant AGN physics. A negative result could tell us
that the so-called ``blazar sequence'' \citep{Sam96, Fos98, Ghi98}
holds to some level \citep{Pad07}, which would mean that the particles
in the relativistic jets of AGN strongly interact with the ambient
photon field, such as the one produced by, e.g., the accretion disk -
BLR complex, already in the most inner regions where they are
produced. A positive result, on the other hand, could mean that the
basic properties of AGN jets, such as, e.g., their powers, magnetic
field strengths or velocities, are not determined by the presence or
absence of emission line regions in the nuclear regions.

Finding radio quasars with synchrotron X-rays in large numbers will
also provide present and up-coming $\gamma$-ray missions, such as
AGILE and the Gamma Ray Large Area Space Telescope (GLAST), with
plenty of targets. A synchrotron spectrum peaking at UV/soft-X-ray
energies produces an inverse Compton spectrum with peak at
$\gamma$-ray frequencies. But such jets have been detected so far only
in a (small) part of the BL Lac population, the HBL. Since at the high
fluxes required by the sensitivity of these missions radio quasars are
by far more abundant than BL Lacs \citep[e.g.,][]{P07}, high-energy
peaked FSRQ instead could become the prime targets.

\section{Summary and Conclusions}

A considerable fraction of FSRQ discovered in the DXRBS and RGB survey
have multiwavelength flux ratios similar to those of BL Lacs with
synchrotron X-rays. However, the definite proof of the synchrotron
nature of their X-ray emission requires X-ray
spectroscopy. \citet{P02} observed four ``X-ray loud'' radio quasars
and found one object (RGB J1629$+$4008) to have X-rays dominated by
synchrotron radiation. In this paper we have presented X-ray
spectroscopy of a further ten of these sources. The main results are:

\vspace*{0.2cm}

1. The X-ray spectrum of 7/10 sources is well fit by a single
power-law with spectral index $\Gamma \la 2$, indicating that
synchrotron X-ray emission is not the dominant component. The
remaining three sources are well fit by a broken power-law with soft
and hard spectral indices $\Gamma_{\rm soft}>2$ and $\Gamma_{\rm
  hard}<2$, respectively, which, as in ``intermediate'' BL Lacs,
suggests that we observe the spectral transition between the
synchrotron and inverse Compton jet components.

2. The lack of success in finding radio quasars with X-ray spectra
dominated by jet synchrotron emission can be attributed to the
employed selection method. This was developed for BL Lacs and requires
the sources to have high both optical/UV and X-ray fluxes relative to
that at radio frequencies (low $\alpha_{\rm rx}$ and $\alpha_{\rm ro}$
values). In the case of radio quasars these criteria yield
predominantly those (few) sources that have a thermal and non-thermal
accretion disk component strong enough to dominate over the (weakly
beamed) jet emission. The majority of the observed FSRQ are either
lobe-dominated (i.e., have log $R<0$; 4/10 objects) or are only
slightly core-dominated (i.e., have $0<\log R<0.3$; 4/10 objects).

3. The discovered radio quasars with their relatively low $\alpha_{\rm
  ro}$ values and strong accretion disk signatures represent a
population intermediate between ``classical'' FSRQ and radio-quiet
quasars. This makes them unique probes for studying the poorly known
accretion disk - jet connection.

4. A refined selection technique based on the work of \citet{L06} is
proposed to be used to search for high-energy peaked FSRQ, which,
since their inverse Compton emission is expected to peak at
$\gamma$-ray frequencies, could be prime targets for GLAST and
AGILE. The recipe is: (i) choose highly core-dominated radio quasars,
and among these (ii) choose radio quasars with low radio core to X-ray
luminosity ratios of log $L_{\rm core}/L_{\rm x} \la 5$.

\acknowledgments 

H.L. acknowledges financial support from the Deutsche Akademie der
Naturforscher Leopoldina (grant no. BMBF-LPD 9901/8-99) and from NASA
(grants no. NNG04GN17G and GO5-6102X). H.L. thanks the European
Southern Observatory and St. John's College, Oxford, where part of
this work was conducted, for their hospitality. We thank Sally
Laurent-Muehleisen for providing the spectra of sources from the RGB
survey in electronic format.

Facilities: \facility{XMM}, \facility{CXO}, \facility{ROSAT},
\facility{GALEX}, \facility{CTIO:2MASS}, \facility{FLWO:2MASS},
\facility{KPNO:2.1m}, \facility{Mayall}, \facility{MMT},
\facility{Sloan}, \facility{VLA}


\clearpage

\begin{deluxetable}{lccccrrrcccc}
\tabletypesize{\small} 
\rotate
\tablecaption{
\label{general} 
General Properties of the Observed Sources} 
\tablewidth{0pt} 
\tablehead{ 
Object Name & R.A.(J2000) & Decl.(J2000) & z & $f_{1keV}$ & $\alpha_{\rm r}$ & $f_{\rm core}$ & log $R$ & 
$\alpha_{\rm ro}$ & $\alpha_{\rm ox}$ & $\alpha_{\rm rx}$ & galactic $N_{\rm H}$ \\ 
&&&& [$\mu$Jy] && [mJy] &&&&& [$10^{20}$ cm$^{-2}$] \\
(1) & (2) & (3) & (4) & (5) & (6) & (7) & (8) & (9) & (10) & (11) & (12)
}
\startdata
RGB J0112$+$3818 & 01 12 18.049 & $+$38 18 56.90 & 0.333 & 0.096 & $ $0.09 &  32.5 & $ $0.65 & 0.38 & 1.38 & 0.72 & 5.64 \\
RGB J0254$+$3931 & 02 54 42.629 & $+$39 31 34.75 & 0.291 & 0.344 & $-$0.31 & 160.2 & $ $0.24 & 0.44 & 1.39 & 0.76 & 9.54 \\
RGB J2229$+$3057 & 22 29 34.151 & $+$30 57 12.10 & 0.322 & 0.203 & $ $0.15 &  44.2 & $-$0.52 & 0.40 & 1.42 & 0.74 & 6.79 \\
RGB J2256$+$2618 & 22 56 39.163 & $+$26 18 43.55 & 0.121 & 0.208 & $-$0.02 &  21.4 & $ $0.19 & 0.43 & 1.19 & 0.69 & 5.20 \\
RGB J2318$+$3048 & 23 18 36.905 & $+$30 48 37.00 & 0.103 & 0.185 & $ $0.12 &  17.5 & $ $0.12 & 0.29 & 1.35 & 0.65 & 6.34 \\
WGA J0110$-$1647               & 01 10 35.516 & $-$16 48 27.80 & 0.781 & 0.176 & $ $0.35 &  72.0 & $ $0.22 & 0.36 & 1.35 & 0.69 & 1.62 \\
WGA J0304$+$0002               & 03 04 58.973 & $+$00 02 35.85 & 0.563 & 0.121 & $ $0.40 &   8.4 & $-$1.22 & 0.50 & 1.21 & 0.74 & 7.00 \\
WGA J0447$-$0322               & 04 47 54.727 & $-$03 22 42.20 & 0.774 & 0.348 & $ $0.47 &  68.0 & $>$2.63 & 0.32 & 1.31 & 0.66 & 4.01 \\
WGA J1026$+$6746               & 10 26 33.850 & $+$67 46 12.10 & 1.181 & 0.069 & $ $0.49 &  84.7 & $-$0.50 & 0.51 & 1.25 & 0.76 & 2.18 \\
WGA J2347$+$0852               & 23 47 38.144 & $+$08 52 46.35 & 0.292 & 0.176 & $ $0.58 &   8.3 & $-$1.30 & 0.38 & 1.34 & 0.71 & 6.02 \\
\enddata

\tablecomments{\footnotesize Columns: (1) object name; (2) and (3)
position of the radio core measured on VLA 1.4 GHz A array map; (4)
redshift; (5) unabsorbed {\it ROSAT} X-ray flux density at 1 keV,
calculated using an X-ray spectral index derived from hardness ratios;
(6) radio spectral index between 1.4 and 5 GHz, calculated from the
sum of the fluxes of all NVSS sources within a $3'$ radius
(corresponding roughly to the beam size of the GB6 survey) and the
total flux from the GB6 and PMN surveys for northern and southern
sources, respectively; (7) radio core flux density measured on VLA 1.4
GHz A array map; (8) radio core dominance parameter $R$ at 1.4 GHz,
defined as $R=L_{\rm core}/L_{\rm ext}$, where $L_{\rm core}$ and
$L_{\rm ext}$ are the core and extended luminosities, respectively;
$k$-corrected rest-frame effective spectral indices between: (9) 5 GHz
and 5000~\AA, (10) 5000~\AA~and 1 keV, and (11) 5 GHz and 1 keV; and
(12) galactic hydrogen column density from \citet{DL90}.}

\end{deluxetable}

\clearpage
\begin{deluxetable}{cccccccccccc}
\tabletypesize{\scriptsize}
\rotate 
\tablecaption{
\label{xmm} 
{\sl XMM-Newton} Journal of Observations} 
\tablewidth{0pt} 
\tablehead{ 
Object Name & \multicolumn{3}{c}{MOS 1} & \multicolumn{3}{c}{MOS 2} & 
\multicolumn{3}{c}{PN} & observation & observation \\
& tot. exp. & filt. exp. & source & tot. exp. & filt. exp. & source & 
tot. exp. & filt. exp. & source & ID & date \\
& [sec] & [sec] & counts & [sec] & [sec] & counts & [sec] & [sec] & counts && \\
(1) & (2) & (3) & (4) & (5) & (6) & (7) & (8) & (9) & (10) & (11) & (12)
}
\startdata
WGA J0110$-$1647 &  7506 &  3106 &  812 &  7508 &  3787 & 1004 &  7242 &  4233 & 5785 
& 0203160401 & 12/17/2003 \\ 
WGA J0304$+$0002 & 15157 & 14638 &  939 & 15162 & 14642 &  928 & 14964 & 10432 & 2733
& 0203160201 & 07/19/2004 \\
WGA J0447$-$0322 &  7657 &  7439 & 1131 &  7662 &  7443 & 1162 &  7464 &  5238 & 3129
& 0203160101 & 03/06/2004 \\
WGA J1026$+$6746 &  9957 &  9571 &  433 &  9962 &  9576 &  412 &  9764 &  6779 & 1420
& 0203160601 & 04/19/2004 \\
\enddata

\tablecomments{\footnotesize Columns: (1) object name; (2) total
exposure time; (3) filtered live exposure time; and (4) extracted
source counts for MOS 1; (5) total exposure time; (6) filtered live
exposure time; and (7) extracted source counts for MOS 2; (8) total
exposure time; (9) filtered live exposure time; and (10) extracted
source counts for PN; (11) observation ID; and (12) observation date.}

\end{deluxetable}
\clearpage

\begin{deluxetable}{lcccccc}
\tabletypesize{\small} 
\tablecaption{
\label{chandra} 
{\sl Chandra} Journal of Observations} 
\tablewidth{0pt} 
\tablehead{ 
\colhead{Object Name} & \colhead{tot. exp.} & \colhead{filt. exp.} & 
\colhead{source} & \colhead{angle} & \colhead{observation} & 
\colhead{observation} \\
 & \colhead{[sec]} & \colhead{[sec]} & \colhead{counts} & 
\colhead{[deg]} & \colhead{ID} & \colhead{date} \\
\colhead{(1)} & \colhead{(2)} & \colhead{(3)} & \colhead{(4)} & 
\colhead{(5)} & \colhead{(6)} & \colhead{(7)}
}
\startdata
RGB J0112$+$3818 & 16183 & 13815 & 1897 & 172 & 5641 & 10/07/2005 \\ 
RGB J0254$+$3931 & 10708 &  8613 & 1584 & 153 & 5638 & 10/26/2005 \\
RGB J2229$+$3057 & 10628 &  9235 & 3884 &  91 & 5639 & 05/04/2005 \\
RGB J2256$+$2618 &  9854 &  8936 & 2042 &  91 & 5642 & 05/04/2005 \\
RGB J2318$+$3048 & 11275 &  9219 & 1168 & 316 & 5643 & 01/25/2005 \\
WGA J2347$+$0852 & 16675 & 14440 & 4936 & 128 & 5640 & 08/20/2005 \\
\enddata

\tablecomments{\footnotesize Columns: (1) object name; (2) total
exposure time; (3) filtered live exposure time; (4) extracted source
counts; (5) nominal roll angle; (6) observation ID; and (7)
observation date.}

\end{deluxetable}


\begin{deluxetable}{lllccll}
\tabletypesize{\small} 
\tablecaption{
\label{epic1} 
EPIC Single Power-Law Fits} 
\tablewidth{0pt}
\tablehead{
Object Name & $N_{\rm H}$ & $\Gamma$ & $f_{(0.2-2.4)}$ & $f_{(2-10)}$ & $\chi^2_{\nu}$/dof & F-test\\
& [$10^{20}$ cm$^{-2}$] & & [erg s$^{-1}$ cm$^{-2}$] & [erg s$^{-1}$ cm$^{-2}$] && null prob.\\
(1) & (2) & (3) & (4) & (5) & (6) & (7)
}
\startdata
WGA J0110$-$1647 & 1.62 fixed             & 1.96$\pm$0.04          & 1.44e$-$12 & 1.19e$-$12 
& 1.08/306 &            \\
                 & 0.62$^{+0.64}_{-0.58}$ & 1.88$^{+0.07}_{-0.06}$ & 1.48e$-$12 & 1.28e$-$12 
& 1.07/305 & 0.011      \\
WGA J0304$+$0002 & 7.00 fixed             & 1.93$\pm$0.05          & 2.98e$-$13 & 3.53e$-$13 
& 0.92/195 &            \\
                 & 4.90$^{+1.59}_{-1.14}$ & 1.83$^{+0.09}_{-0.06}$ & 2.96e$-$13 & 3.75e$-$13 
& 0.90/194 & 0.019      \\
WGA J0447$-$0322 & 4.01 fixed             & 2.11$\pm$0.04          & 7.31e$-$13 & 5.63e$-$13 
& 1.11/219 &            \\
                 & 1.79$^{+0.84}_{-0.72}$ & 1.97$^{+0.07}_{-0.06}$ & 7.60e$-$13 & 6.34e$-$13 
& 1.04/218 & 7.10e$-$05 \\
WGA J1026$+$6746 & 2.18 fixed             & 1.76$\pm$0.08          & 1.99e$-$13 & 2.39e$-$13 
& 1.06/97  &            \\
                 & 0.60$^{+1.84}_{-0.60}$ & 1.68$^{+0.13}_{-0.09}$ & 2.08e$-$13 & 2.55e$-$13 
& 1.05/96  & 0.171      \\
\enddata

\tablecomments{\footnotesize Columns: (1) object name; (2) hydrogen
  column density; (3) photon index; (4) observed PN flux in the range
  0.2-2.4 keV; (5) observed PN flux in the 2-10 keV range; (6) reduced
  $\chi^2$ and number of degrees of freedom; and (7) $F$-test null
  probability quantifying the significance of the $\chi^2$ decrease
  due to the addition of a new parameter (free $N_{\rm H}$). The
  errors are quoted at 90\% confidence.}

\end{deluxetable}

\begin{deluxetable}{lllccll}
\tabletypesize{\small} 
\tablecaption{
\label{acis1} 
ACIS Single Power-Law Fits} 
\tablewidth{0pt}
\tablehead{
Object Name & $N_{\rm H}$ & $\Gamma$ & $f_{(0.3-2.4)}$ & $f_{(2-10)}$ & $\chi^2_{\nu}$/dof & F-test\\
& [$10^{20}$ cm$^{-2}$] & & [erg s$^{-1}$ cm$^{-2}$] & [erg s$^{-1}$ cm$^{-2}$] && null prob.\\
(1) & (2) & (3) & (4) & (5) & (6) & (7)
}
\startdata
RGB J0112$+$3818 & 5.64 fixed                       & 1.89$\pm$0.08          & 4.49e$-$13 & 5.44e$-$13 & 1.07/69  &            \\
                 & 4.36e$-$08$^{+1.97}_{-4.36e-08}$ & 1.67$^{+0.09}_{-0.06}$ & 4.77e$-$13 & 6.25e$-$13 & 0.88/68  & 2.32e$-$04 \\
RGB J0254$+$3931 & 9.54 fixed                       & 1.80$\pm$0.08          & 5.74e$-$13 & 8.99e$-$13 & 0.83/62  &            \\
                 & 6.85$^{+3.51}_{-3.27}$           & 1.71$^{+0.14}_{-0.13}$ & 5.80e$-$13 & 9.47e$-$13 & 0.81/61  & 0.163      \\
RGB J2229$+$3057 & 6.79 fixed                       & 2.04$\pm$0.05          & 1.39e$-$12 & 1.40e$-$12 & 1.31/128 &            \\
                 & 4.84e$-$07$^{+1.34}_{-4.84e-07}$ & 1.76$^{+0.07}_{-0.04}$ & 1.48e$-$12 & 1.71e$-$12 & 0.98/127 & 8.69e$-$10 \\
RGB J2256$+$2618 & 5.20 fixed                       & 1.66$\pm$0.06          & 7.12e$-$13 & 1.18e$-$12 & 1.07/77  &            \\
                 & 4.85$^{+2.84}_{-2.66}$           & 1.65$^{+0.12}_{-0.11}$ & 7.13e$-$13 & 1.19e$-$12 & 1.08/76  & 0.789      \\
RGB J2318$+$3048 & 6.34 fixed                       & 1.67$\pm$0.09          & 3.93e$-$13 & 6.70e$-$13 & 0.78/46  &            \\
                 & 1.91$^{+4.01}_{-1.91}$           & 1.52$^{+0.16}_{-0.13}$ & 4.04e$-$13 & 7.38e$-$13 & 0.72/45  & 0.039      \\
WGA J2347$+$0852 & 6.02 fixed                       & 1.86$\pm$0.04          & 1.10e$-$12 & 1.40e$-$12 & 1.34/147 &            \\
                 & 3.83$^{+1.43}_{-1.35}$           & 1.78$\pm$0.07          & 1.12e$-$12 & 1.49e$-$12 & 1.30/146 & 0.031      \\
\enddata

\tablecomments{\footnotesize Columns: (1) object name; (2) hydrogen
  column density; (3) photon index; (4) observed ACIS flux in the
  range 0.3-2.4 keV; (5) observed ACIS flux in the 2-10 keV range; (6)
  reduced $\chi^2$ and number of degrees of freedom; and (7) $F$-test
  null probability quantifying the significance of the $\chi^2$
  decrease due to the addition of a new parameter (free $N_{\rm
  H}$). The errors are quoted at 90\% confidence.}

\end{deluxetable}


\begin{deluxetable}{llllcccll}
\tabletypesize{\footnotesize}
\rotate
\tablecaption{
\label{epic2} 
EPIC Broken Power-Law Fits} 
\tablewidth{0pt} 
\tablehead{ 
Object Name & $N_{\rm H}$ & $\Gamma_{\rm soft}$ & $\Gamma_{\rm hard}$ 
& $E_{\rm break}$ & $f_{(0.2-2.4)}$ & $f_{(2-10)}$ & $\chi^2_{\nu}$/dof & F-test \\
& [$10^{20}$ cm$^{-2}$] & & & [keV] & [erg s$^{-1}$ cm$^{-2}$] & [erg s$^{-1}$ cm$^{-2}$] && null prob. \\
(1) & (2) & (3) & (4) & (5) & (6) & (7) & (8) & (9)
}
\startdata
WGA J0110$-$1647 & 1.62 fixed             & 1.97$\pm$0.04          
& $-$3.00(unc.)          & 6.91$^{+0.47}_{-0.39}$ & 1.43e$-$12 & 1.68e$-$12 
& 1.03/304 & 1.29e$-$04 \\
WGA J0304$+$0002 & 7.00 fixed             & 2.07$^{+0.33}_{-0.10}$ 
& 1.81$^{+0.10}_{-0.09}$ & 1.26$^{+0.54}_{-0.55}$ & 2.92e$-$13 & 3.78e$-$13
& 0.89/193 & 0.012 \\
WGA J0447$-$0322 & 4.01 fixed             & 2.28$^{+0.07}_{-0.08}$ 
& 1.74$^{+0.11}_{-0.17}$ & 1.62$^{+0.53}_{-0.32}$ & 7.26e$-$13 & 7.27e$-$13 
& 0.94/217 & 9.71e$-$09 \\
WGA J1026$+$6746 & 2.18 fixed             & 1.96$^{+0.82}_{-0.23}$ 
& 1.66$^{+0.11}_{-0.27}$ & 0.99$^{+2.05}_{-0.54}$ & 2.02e$-$13 & 2.61e$-$13 
& 1.05/95 & 0.246 \\
\enddata

\tablecomments{Columns: (1) object name; (2) hydrogen column density;
  (3) soft photon index; (4) hard photon index; (5) break energy; (6)
  observed PN flux in the range 0.2-2.4 keV; (7) observed PN flux in
  the 2-10 keV range; (8) reduced $\chi^2$ and number of degrees of
  freedom and (9) $F$-test null probability quantifying the
  significance of the $\chi^2$ decrease due to the addition of two
  parameters (from a single power-law fit with Galactic absorption to
  a broken power-law fit). The errors are quoted at 90\% confidence.}

\end{deluxetable}

\begin{deluxetable}{llllcccll}
\tabletypesize{\footnotesize}
\rotate
\tablecaption{
\label{acis2} 
ACIS Broken Power-Law Fits} 
\tablewidth{0pt} 
\tablehead{ 
Object Name & $N_{\rm H}$ & $\Gamma_{\rm soft}$ & $\Gamma_{\rm hard}$ 
& $E_{\rm break}$ & $f_{(0.3-2.4)}$ & $f_{(2-10)}$ & $\chi^2_{\nu}$/dof & F-test \\
& [$10^{20}$ cm$^{-2}$] & & & [keV] & [erg s$^{-1}$ cm$^{-2}$] & [erg s$^{-1}$ cm$^{-2}$] && null prob. \\
(1) & (2) & (3) & (4) & (5) & (6) & (7) & (8) & (9)
}
\startdata
RGB J0112$+$3818 & 5.64 fixed & 2.57$^{+0.59}_{-0.29}$ & 1.74$\pm$0.09          & 0.86$^{+0.15}_{-0.17}$ & 4.72e$-$13 & 6.18e$-$13 & 0.82/67  & 6.89e$-$05 \\
RGB J0254$+$3931 & 9.54 fixed & 1.97$^{+0.55}_{-0.27}$ & 1.72$^{+0.11}_{-0.84}$ & 1.17(unc.)             & 5.76e$-$13 & 9.50e$-$13 & 0.81/60  & 0.223      \\
RGB J2229$+$3057 & 6.79 fixed & 2.31$^{+0.22}_{-0.11}$ & 1.70$^{+0.17}_{-0.15}$ & 1.50$^{+0.43}_{-0.47}$ & 1.40e$-$12 & 1.78e$-$12 & 1.02/126 & 3.79e$-$08 \\
RGB J2256$+$2618 & 5.20 fixed & 1.69$^{+2.88}_{-4.69}$ & 1.59$^{+8.41}_{-4.59}$ & 2.34(unc.)             & 7.10e$-$13 & 1.22e$-$12 & 1.09/75  & 0.775      \\
RGB J2318$+$3048 & 6.34 fixed & 1.78$^{+0.09}_{-0.12}$ & 1.00$^{+0.51}_{-1.30}$ & 3.01$^{+1.76}_{-1.49}$ & 3.91e$-$13 & 8.72e$-$13 & 0.65/44  & 0.007      \\
WGA J2347$+$0852 & 6.02 fixed & 1.90$\pm$0.05          & 0.60$^{+0.74}_{-1.48}$ & 4.64$^{+1.17}_{-0.97}$ & 1.10e$-$12 & 1.82e$-$12 & 1.28/145 & 0.013      \\
\enddata

\tablecomments{Columns: (1) object name; (2) hydrogen column density;
  (3) soft photon index; (4) hard photon index; (5) break energy; (6)
  observed ACIS flux in the range 0.3-2.4 keV; (7) observed ACIS flux
  in the 2-10 keV range; (8) reduced $\chi^2$ and number of degrees of
  freedom and (9) $F$-test null probability quantifying the
  significance of the $\chi^2$ decrease due to the addition of two
  parameters (from a single power-law fit with Galactic absorption to
  a broken power-law fit).  The errors are quoted at 90\% confidence.}

\end{deluxetable}


\begin{deluxetable}{lccc|cc|cc}
\tabletypesize{\small}
\tablecaption{
\label{mag} 
Near-IR, Optical and UV Magnitudes} 
\tablewidth{0pt} 
\tablehead{ 
Object Name & \multicolumn{3}{c|}{2MASS\tablenotemark{\star}} & 
\multicolumn{2}{c|}{APM} & \multicolumn{2}{c}{GALEX} \\
& J & H & Ks & O & E & FUV & NUV \\ 
& 1.235 $\mu$m & 1.622 $\mu$m & 2.159 $\mu$m & 4100~\AA & 6500~\AA & 1530~\AA & 2310~\AA \\ 
& [mag] & [mag] & [mag] & [mag] & [mag] & [mag] & [mag] \\
(1) & (2) & (3) & (4) & (5) & (6) & (7) & (8)
}
\startdata
RGB J0112$+$3818 & 15.5 & 14.6 & 13.9 & 17.3 & 16.5 &      &      \\
RGB J0254$+$3931 & 15.9 (16.1) & 15.3 (15.4) & 14.4 (13.9) & 
                   16.4 & 15.4 &      &      \\
RGB J2229$+$3057 & 15.5 & 14.7 & 14.3 & 16.4 & 15.8 &      &      \\
RGB J2256$+$2618 & 15.9 (15.0) & 15.1 (13.8) & 14.6 (13.8) & 
                   18.3 & 16.7 & 21.7 & 21.0 \\
RGB J2318$+$3048 & 15.2 (14.3) & 14.4 (13.9) & 13.9 (13.3) & 
                   17.3 &      & 20.2 & 19.5 \\
WGA J0110$-$1647 & 15.0 & 14.7 & 14.0 & 15.9 & 16.2 & 17.2 & 16.5 \\
WGA J0304$+$0002 & 16.5 & 15.9 & 15.2 &      & 18.4 & 19.7 & 19.1 \\
WGA J0447$-$0322 & 14.7 & 14.5 & 13.8 & 16.3 & 16.0 &      &      \\
WGA J2347$+$0852 & 15.6 (15.2) & 14.8 (14.2) & 13.8 (13.5) & 
                   17.1 & 16.2 & 18.0 & 17.7 \\
\enddata

\tablenotetext{\star}{magnitudes from the Extended Source catalog are
listed in parenthesis}

\end{deluxetable}

\begin{deluxetable}{lcccccc}
\tabletypesize{\small} 
\tablecaption{
\label{om} 
{\sl XMM-Newton} Optical Monitor Data} 
\tablewidth{0pt} 
\tablehead{ 
Object Name & V & B & U & UVW1 & UVM2 & observation \\
& $\lambda5430$ & $\lambda4500$ & $\lambda3440$ 
& $\lambda2910$ & $\lambda2310$ & date \\
& [mag] & [mag] & [mag] & [mag] & [mag] & \\
(1) & (2) & (3) & (4) & (5) & (6) & (7)
}
\startdata
WGA J0110$-$1647 & 16.42$\pm$0.01 & 16.642$\pm$0.007 & 15.587$\pm$0.006 
& 15.241$\pm$0.008 &              & 12/17/2003 \\ 
WGA J0304$+$0002 & 18.76$\pm$0.02 & 19.04$\pm$0.07   & 18.07$\pm$0.05
& 17.75$\pm$0.05 &                & 07/19/2004 \\
WGA J0447$-$0322 & 16.33$\pm$0.03 & 16.66$\pm$0.01   & 15.76$\pm$0.01
&                &                & 03/06/2004 \\
WGA J1026$+$6746 & 18.41$\pm$0.06 &                  & 17.49$\pm$0.02
& 17.07$\pm$0.02 & 17.23$\pm$0.05 & 03/23/2004 \\
                 &                &                  & 
& 17.09$\pm$0.03 &                & 04/19/2004 \\
\enddata
\end{deluxetable}


\begin{deluxetable}{lcccccc}
\tabletypesize{\small}
\tablecaption{
\label{acctable} 
Accretion Disk Model Parameters} 
\tablewidth{0pt} 
\tablehead{
Object Name & $L_{\rm BLR}$ & $\dot{M}$ & FWHM & broad 
& $M_{\rm BH}$ & $\dot{m}$ \\
& [erg/s] & [$M_{\odot}$/yr] & [km/s] & line & [$M_{\odot}$] & \\ 
(1) & (2) & (3) & (4) & (5) & (6) & (7)
}
\startdata
RGB J0112$+$3818 & 2.7e$+$44 &  0.8  & 3144 & \MgII    & 3e$+$08 & 0.08  \\
RGB J0254$+$3931 & 1.4e$+$44 &  0.4  & 3244 & H$\beta$ & 2e$+$08 & 0.05  \\
RGB J2229$+$3057 & 2.6e$+$44 &  0.8  & 5467 & H$\beta$ & 8e$+$08 & 0.03  \\
RGB J2256$+$2618 & 6.9e$+$42 &  0.02 & 7888 & H$\beta$ & 3e$+$08 & 0.002 \\ 
RGB J2318$+$3048 & 5.9e$+$42 &  0.02 & 6645 & H$\beta$ & 2e$+$08 & 0.003 \\
WGA J0110$-$1647 & 1.5e$+$45 &  5    & 3853 & \MgII    & 1e$+$09 & 0.1   \\
WGA J0304$+$0002 & 2.7e$+$44 &  0.8  & 7404 & H$\beta$ & 2e$+$09 & 0.01  \\
WGA J0447$-$0322 & 6.5e$+$45 & 19    & 4595 & \MgII    & 3e$+$09 & 0.2   \\
WGA J1026$+$6746 & 1.1e$+$46 & 32    & 6646 & \MgII    & 8e$+$09 & 0.1   \\
WGA J2347$+$0852 & 8.1e$+$43 &  0.2  & 7926 & H$\beta$ & 9e$+$08 & 0.005 \\
\enddata

\tablecomments{\footnotesize Columns: (1) object name, (2) BLR
  luminosity, (3) accretion rate (in solar masses per year) calculated
  from the BLR luminosity assuming a covering factor of 10\%, (4) FWHM
  of the broad emission line listed in (5), (6) mass of the black hole
  (in solar masses) calculated from the FWHM and the BLR luminosity
  assuming a covering factor of 10\%, and (7) dimensionless accretion
  rate defined as $m=\dot{M}/\dot{M_{\rm E}}$, where $\dot{M_{\rm E}}$
  is the accretion rate at the Eddington limit. See text for more
  details.}

\end{deluxetable}

\clearpage

\begin{figure*}
\centerline{
\includegraphics[scale=0.42]{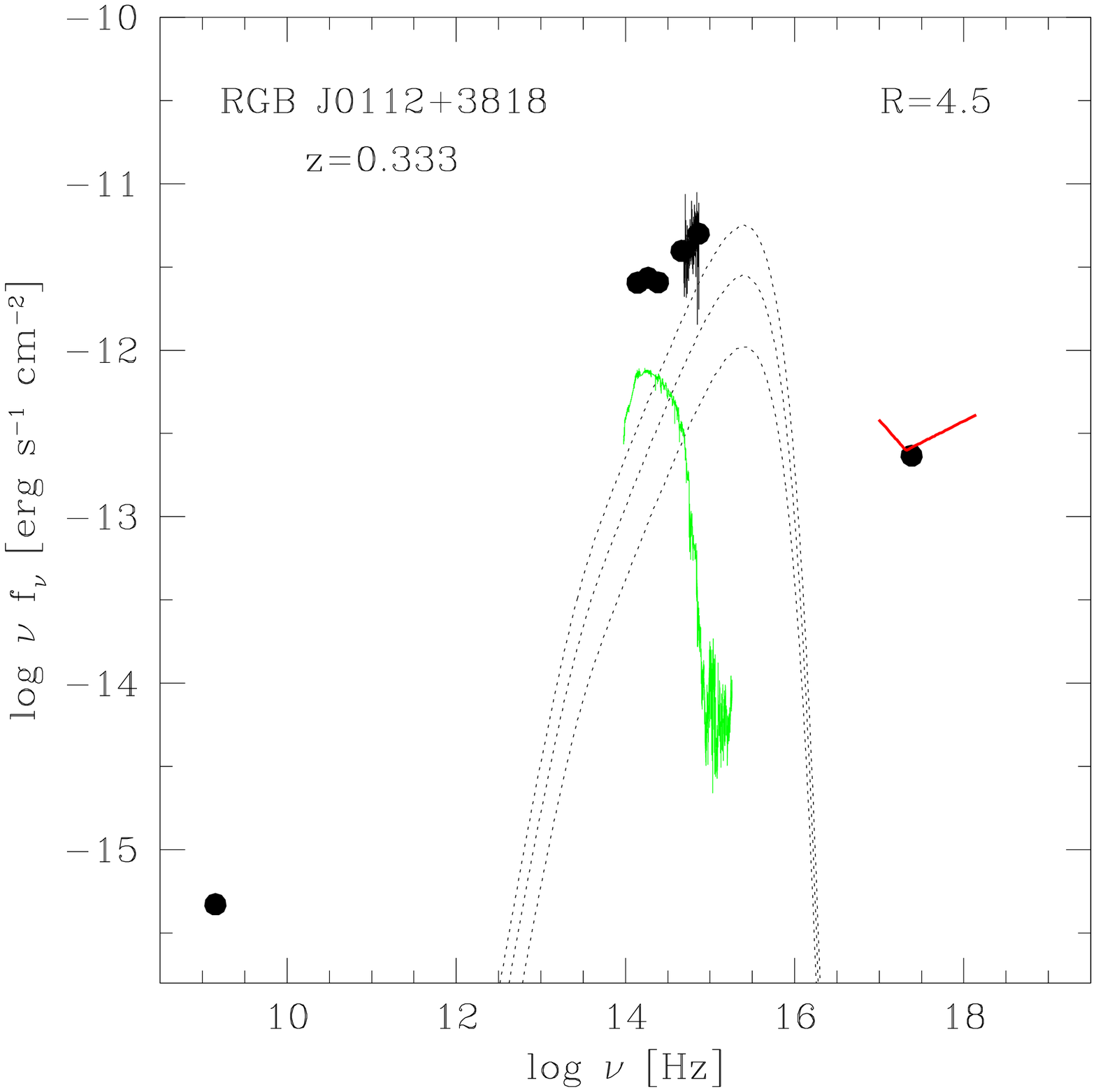}
\includegraphics[scale=0.42]{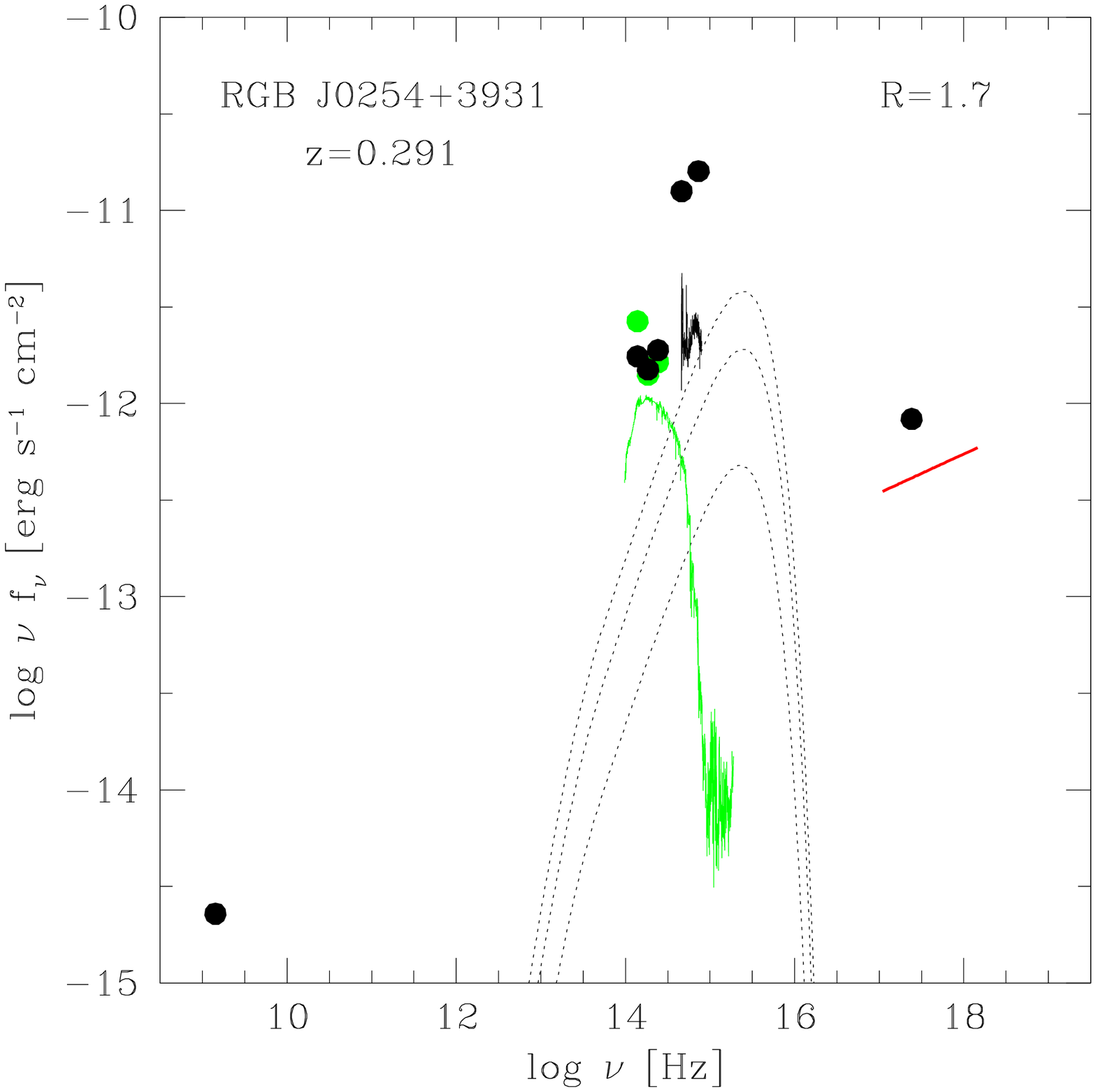}
}
\centerline{
\includegraphics[scale=0.42]{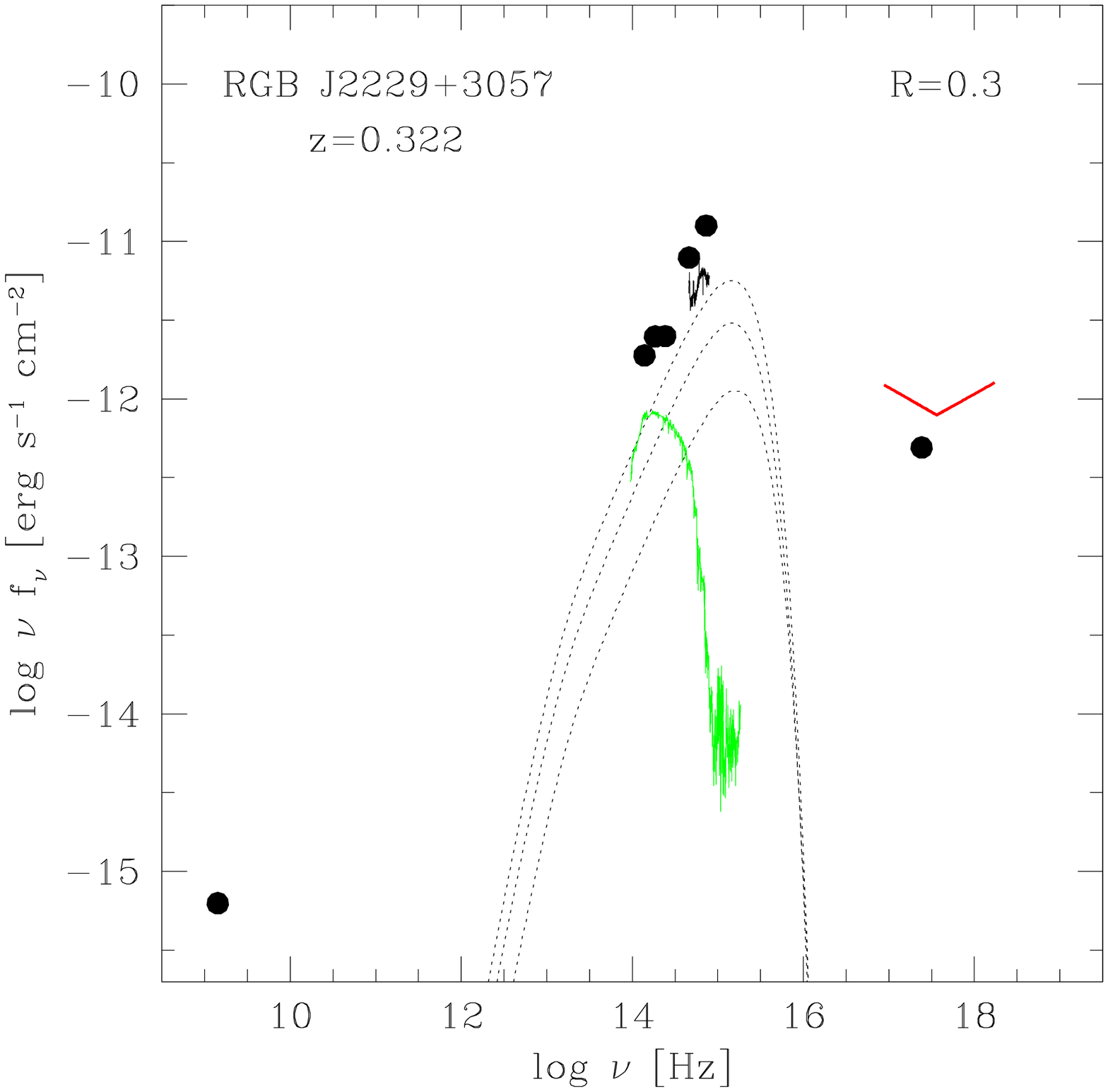}
\includegraphics[scale=0.42]{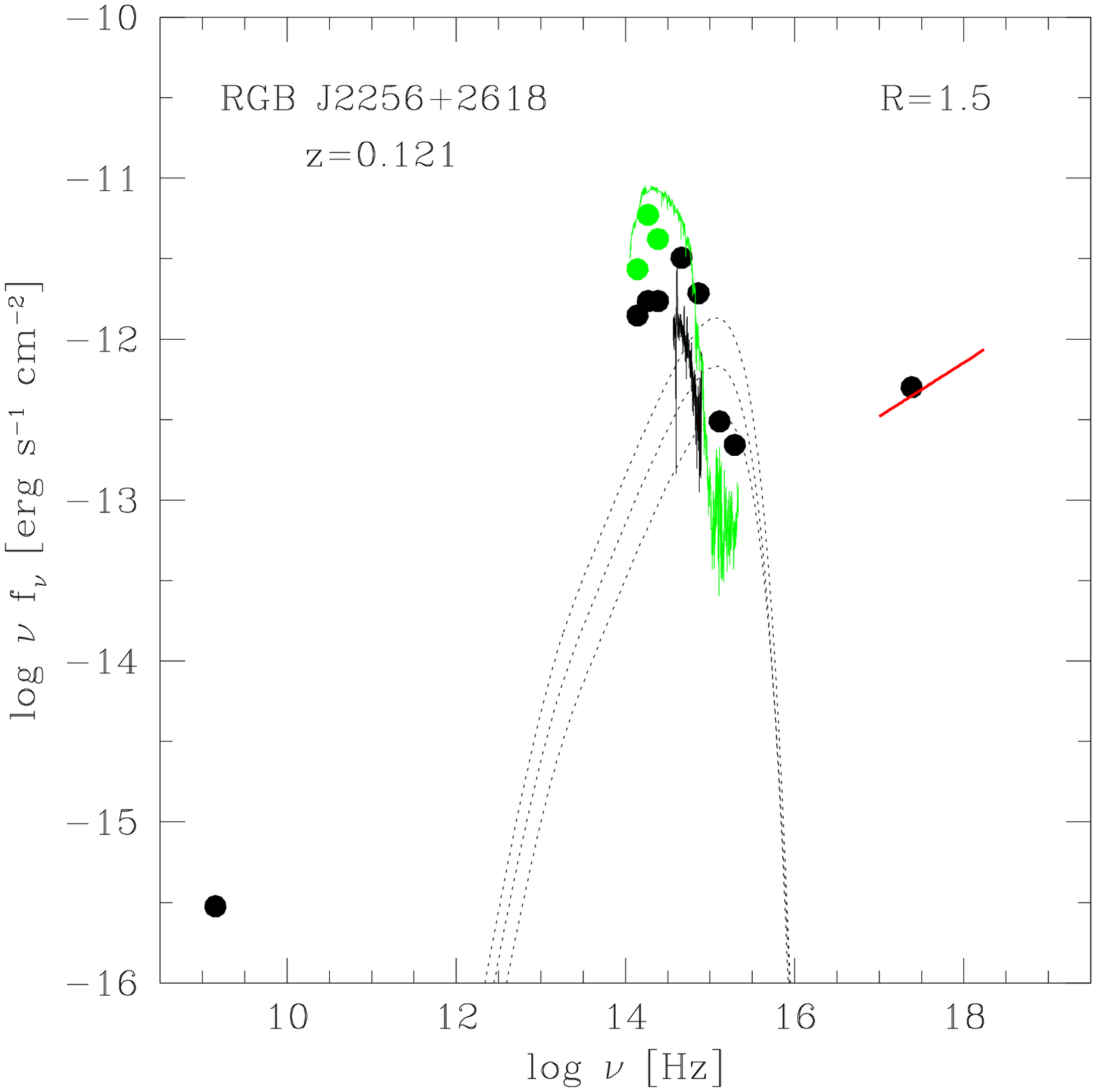}
}
\caption{\label{chandrased} As in Fig. \ref{xmmsed} for the
{\sl Chandra} sources.}
\end{figure*}

\begin{figure*}
\figurenum{\ref{chandrased}}
\centerline{
\includegraphics[scale=0.42]{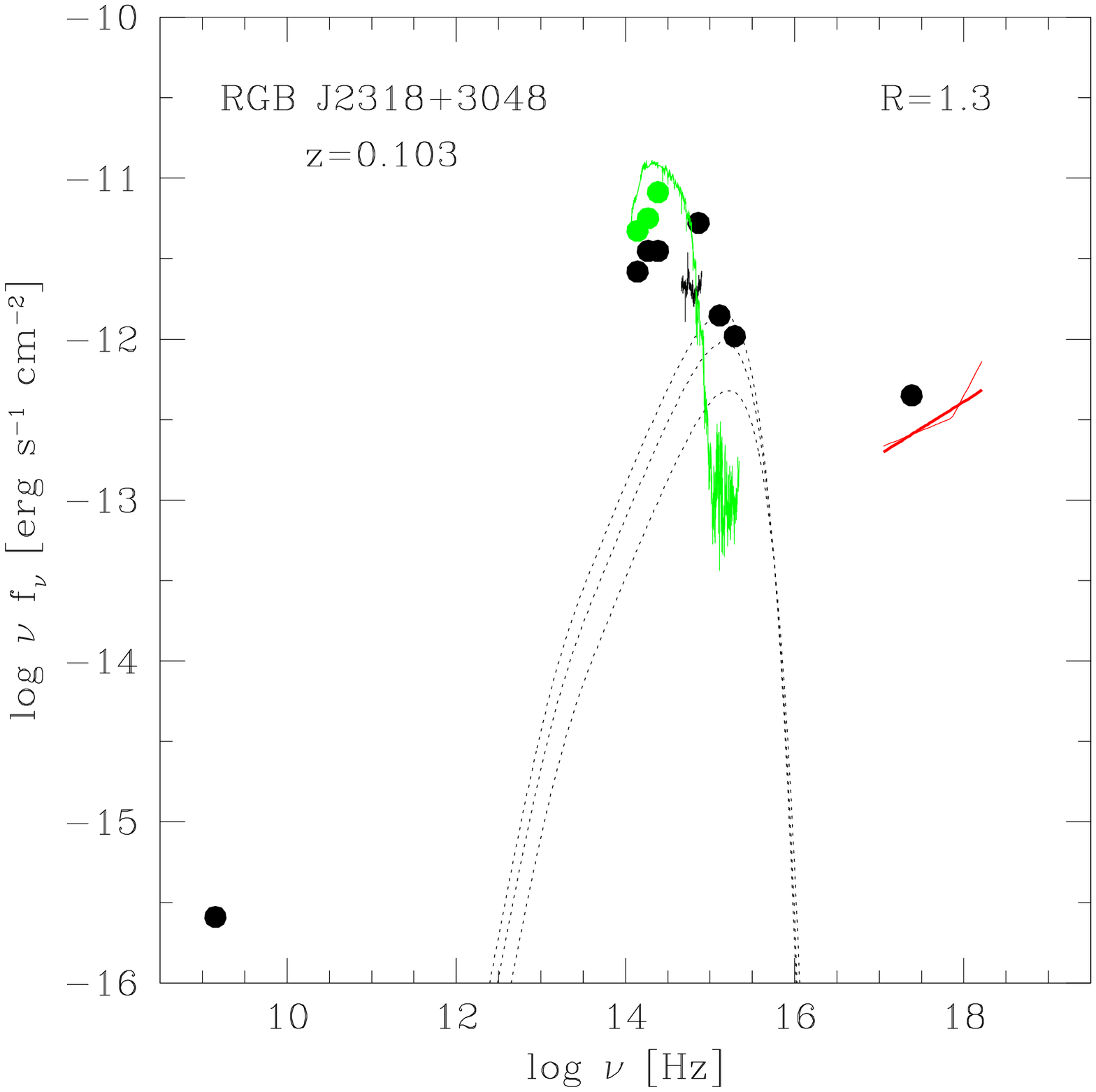}
\includegraphics[scale=0.42]{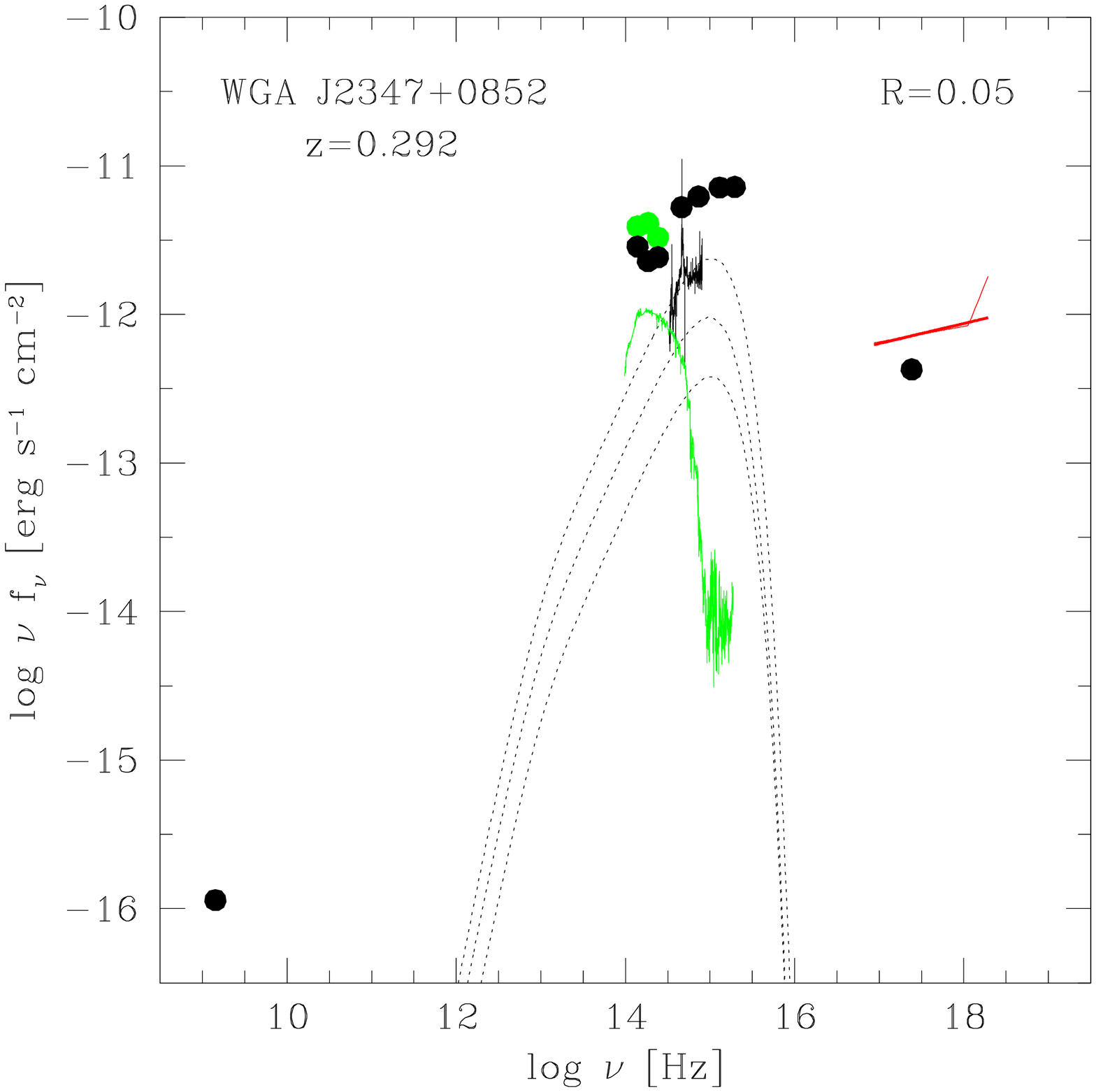}
}
\caption{continued.}
\end{figure*}

\clearpage

\begin{figure*}
\centerline{
\includegraphics[scale=0.42]{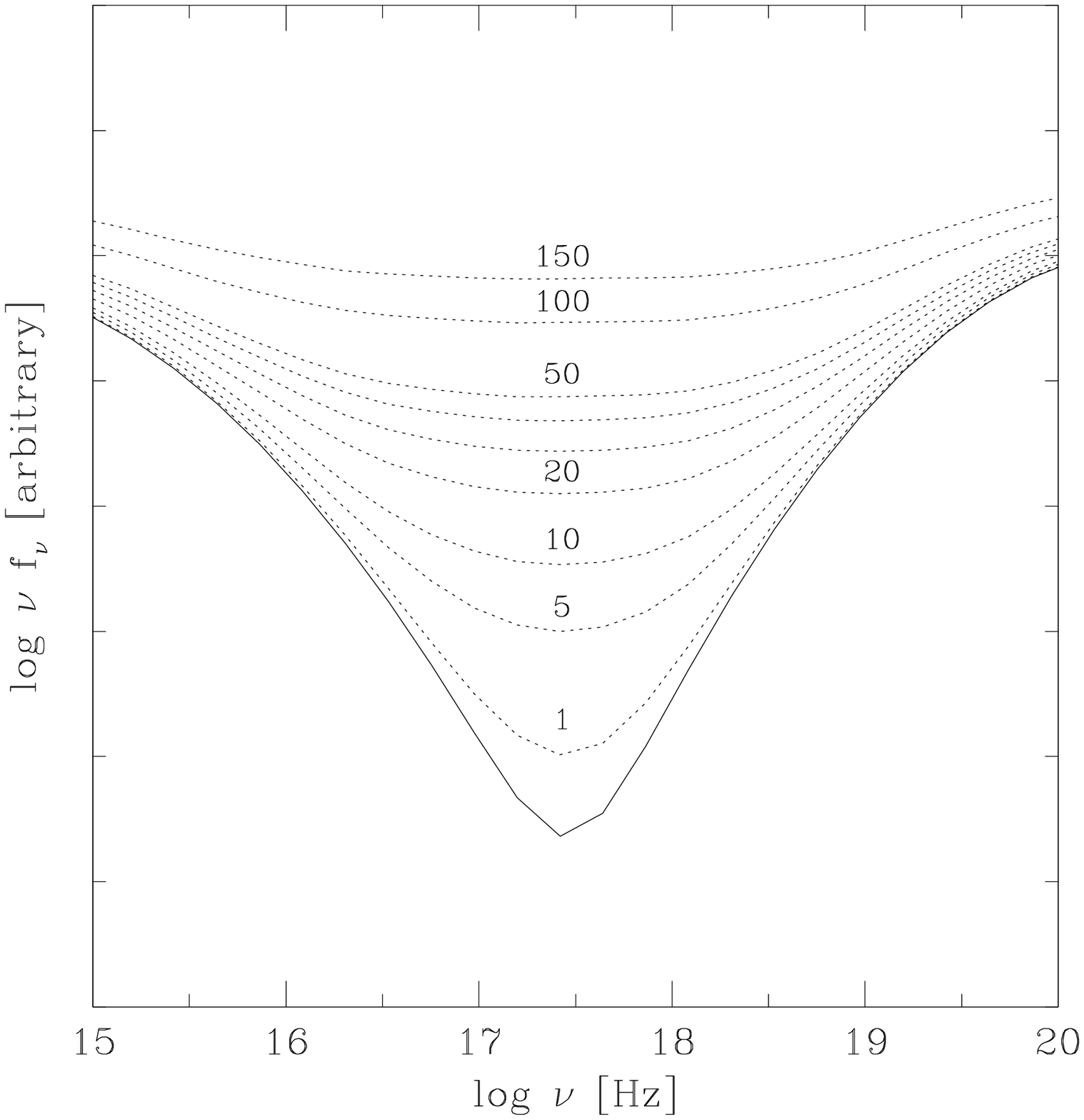}
}
\caption{\label{corona} Simulated change in observed jet SED
around the transition point between the synchrotron and inverse
Compton components as a power-law component with an X-ray spectral
index of $\Gamma \sim 2$ adds to it. Flux ratios of 1, 5, 10, 20, 30,
40, 50, 100, and 150 were assumed between this component and the jet
SED at the transition point.}
\end{figure*}

\end{document}